\documentstyle[12pt]{article}
\begin{document}
\begin{flushright}
MRC--PH--TH.16--94
\end{flushright}
\begin{center}
\Large {\bf Anomaly and Exotic Statistics\\
       in One Dimension}
\vspace{0.25in}

\large

\bigskip

Fuad M. Saradzhev\footnote{E-mail: fuad@yunus.mam.tubitak.gov.tr}\\
{\it Institute of Physics, Academy of Sciences of Azerbaijan,\\
\it Huseyn Javid pr. 33, 370143 Baku,
AZERBAIJAN\footnote{Permanent address}}\\
and\\
{\it Department of Physics, Marmara Research Centre,\\
\it TUBITAK, P.O. Box 21, 41470 Gebze, TURKEY\footnote{Present address}}

\normalsize

\vspace{0.3in}

{ABSTRACT}

\end{center}

\begin{quotation}

We study the influence of the anomaly on the physical quantum
picture of the chiral Schwinger model (CSM) defined on $S^1$.
We show that such phenomena as the total screening of charges
and the dynamical mass generation characteristic for the Schwinger
model do not take place here. Instead of them, the anomaly results
in the background linearly rising electric field or, equivalently,
in the exotic statistics of the physical matter field. We construct
the algebra  of the Poincare generators and show that it differs
from the Poincare one. For the CSM on $R^1$, the anomaly influences
only the mass generation mechanism.

\end{quotation}

\newpage
\section{INTRODUCTION}
\label{sec: intro}
\rm

The two-dimensional QED with massless fermions, i.e. the Schwinger
model (SM), demonstrates such phenomena as the dynamical mass
generation and the total screening of the charge \cite{schw63} .
Although the Lagrangian of the SM contains only massless fields, a
massive boson field emerges out of the interplay of the dynamics
that govern the original fields. This mass generation is due to the
complete compensation of any charge inserted into the vacuum.

In the chiral Schwinger model \cite{jack85,raja85}
only the right (or left) chiral component of the fermionic field
is coupled to the $U(1)$ gauge field. The left-right asymmetric
matter content leads to an anomaly. At the quantum level, the local
gauge symmetry is not realized by a unitary action of the gauge
symmetry group on Hilbert space. The Hilbert space furnishes a
projective representation of the symmetry group
\cite{wign39,jack83,nels85}.

In this paper, we aim to study the influence of the anomaly on the
physical quantum picture of the CSM. Do the dynamical mass generation
and the total screening of charges take place also in the CSM? Are
there any new physical effects caused just by the left-right asymmetry?
These are the questions which we want to answer.

To get the physical quantum picture of the CSM we need first to
construct a self-consistent quantum theory of the model and then
solve all the quantum constraints. In the quantization procedure,
the anomaly manifests itself through a special Schwinger term in the
commutator algebra of the Gauss law generators. This term changes
the nature of the Gauss law constraint: instead of being first-class
constraint, it turns into second-class one. As a consequence, the
physical quantum states cannot be defined as annihilated by the
Gauss law generator.

There are different approaches to overcome this problem and to
consistently quantize the CSM. The fact that the second class
constraint appears only after quantization means that the number
of degrees of freedom of the quantum theory is larger than that of
the classical theory. To keep the Gauss law constraint first-class,
Faddeev and Shatashvili proposed adding an auxiliary field in such a
way that the dynamical content of the model does not change
\cite{fadd86}. At the same time, after quantization it is the auxiliary
field that furnishes the additional "irrelevant" quantum degrees of
freedom. The auxiliary field is described by the Wess-Zumino term.
When this term is added to the Lagrangian of the original model,
a new, anomaly-free model is obtained. Subsequent canonical
quantization of the new model is achieved by the Dirac procedure.

For the CSM, the correspondig WZ-term is not defined uniquely.
It contains the so called Jackiw-Rajaraman parameter $a > 1$.
This parameter reflects an ambiguity in the bosonization procedure
and in the construction of the WZ-term. Although the spectrum of
the new, anomaly-free model turns out to be relativistic and
contains a relativistic boson, the mass of the boson also depends
on the Jackiw-Rajaraman parameter \cite{jack85,raja85}. This
mass is definetely unphysical and corresponds to the unphysical
degrees of freedom. The quantum theory containing such a parameter
in the spectrum is not consistent or, at least, is not that final
version of the quantum theory which we would like to get.

In another approach also formulated by Faddeev \cite{fadd84},
the auxiliary field is not added, so the quantum Gauss law constraint
remains second-class. The standard Gauss law is assumed to be regained
as a statement valid in matrix elements between some states of the
total Hilbert space, and it is the states that are called physical.
The theory is regularized in such a way that the quantum Hamiltonian
commutes with the nonmodified, i.e. second-class quantum Gauss law
constraint. The spectrum is non-relativistic \cite{hall86,para88}.

Here, we follow the approach given in our previous work
\cite{sarad91,sarad93}. The pecularity of the CSM is that its anomalous
behaviour is trivial in the sense that the second class constraint
which appears after quantization can be turned into first class by
a simple redefinition of the canonical variables. This allows us
to formulate a modified Gauss law to constrain physical states.
The physical states are gauge-invariant up to a phase.
In \cite{niemi85,niemi86,semen87}, the modification of the Gauss
law constraint is obtained by making use of the adiabatic approach.

Contrary to \cite{sarad91,sarad93} where the CSM is defined
on $R^1$ , we suppose here that space is a circle of length $L$,
$-\frac{L}{2} \leq x < \frac{L}{2}$ , so space-time manifold is a cylinder
$S^1 \times R^1$ . The gauge field then acquires a global physical
degree of freedom represented by the non-integrable phase of the
Wilson integral on $S^1$. We show that this brings in the physical
quantum picture new features of principle.

Another way of making two-dimensional gauge field dynamics nontrivial
is by fixing the spatial asymptotics of the gauge field
\cite{sarad92,sarad94}. If we assume that the gauge field defined on
$R^1$ diminishes rather rapidly at spatial infinities, then it again
acquires a global physical degree of freedom. We will see that the
physical quantum picture for the model defined on $S^1$ is equivalent
to that obtained in \cite{sarad92,sarad94}.

We work in the temporal gauge $A_0=0$ in the framework of the
canonical quantization scheme and use the Dirac's quantization method
for the constrained systems \cite{dirac64}. In Section~\ref{sec:
quant}, we quantize our model in two steps. First, the matter fields
are quantized, while $A_1$ is handled as a classical background field.
The gauge field $A_1$ is quantized afterwords, using the functional
Schrodinger representation. We derive the anomalous commutators with
nonvanishing Schwinger terms which indicate that our model is
anomalous.

In Section~\ref{sec: const}, we show that the Schwinger term in the
commutator of the Gauss law generators is removed by a redefinition
of these generators and formulate the modified quantum Gauss law
constraint. We prove that this constraint can be also obtained by
using the adiabatic approximation and the notion of quantum holonomy.

In Section~\ref{sec: exoti}, we construct the physical quantum
Hamiltonian consistent with the modified quantum Gauss law constraint,
i.e. invariant under the modified gauge transformations both topologically
trivial and non-trivial. We introduce the modified topologically non-trivial
gauge transformation operator and define $\theta$--states which are its
eigenstates. We define the exotic statistics matter field and reformulate
the quantum theory in terms of this field.

In Section~\ref{sec: poinc}, we construct two other Poincare generators,
i.e. the momentum and the boost. We act in the same way as before with
the Hamiltonian, namely we define the physical generators as those
which are invariant under both topologically trivial and non-trivial
gauge transformations. We show that the algebra of the constructed
generators is not a Poincare one.

In Section~\ref{sec: scree}, we study the charge screening. We introduce
external charges and calculate $(i)$ the energy of the ground state
of the physical Hamiltonian with the external charges and $(ii)$ the
current density induced by these charges.

Section~\ref{sec: discu} contains our conclusions and discussion.

\newpage
\section{QUANTIZATION PROCEDURE}
\label{sec: quant}
\subsection{CLASSICAL THEORY}

The Lagrangian density of the CSM is
\begin{equation}
{\cal L} = - {\frac{1}{4}} {\rm F}_{\mu \nu} {\rm F}^{\mu \nu} +
\bar{\psi} i {\hbar} {\gamma}^{\mu} {\partial}_{\mu} \psi +
e {\hbar} \bar{\psi}_{\rm R} {\gamma}^{\mu} {\psi_{\rm R}} A_{\mu} ,
\label{eq: odin}
\end{equation}
where ${\rm F}_{\mu \nu}= \partial_{\mu} A_{\nu} - \partial_{\nu} A_{\mu}$ ,
$(\mu, \nu) = \overline{0,1}$ , $\gamma^{\mu}$ are $(2 \times 2)$--Dirac
matrices:
\[ \gamma^0 = \left( \begin{array}{cc}
                       0 & 1 \\
                       1 & 0
                     \end{array} \right),
\gamma^1 = \left( \begin{array}{cc}
                    0 & -1 \\
                    1 &  0
                  \end{array} \right),
\gamma^5 = {\gamma^0}{\gamma^1} = \left( \begin{array}{cc}
                                           1 &  0 \\
                                           0 & -1
                                         \end{array} \right) . \]
The field $\psi$ is $2$--component Dirac spinor, $\bar{\psi} =
\psi^{\star} \gamma^0$ and $\psi_{\rm R}=\frac{1}{2} (1+\gamma^5) \psi$.

In the temporal gauge $A_0=0$, the Hamiltonian density without the
left-handed matter part is
\[
{\cal H}_{\rm R}  =  {\cal H}_{\rm EM} + {\cal H}_{\rm F},
\]
\begin{equation}
{\cal H}_{\rm EM}  =  \frac{1}{2} {\rm E}^2,
\label{eq: ddva}
\end{equation}
\[
{\cal H}_{\rm F}  \equiv  \hbar \psi_{\rm R}^{\star} d \psi_{\rm R} =
\hbar\psi_{\rm R}^{\star}(-i{\partial}_{1}-eA_1)\psi_{\rm R},
\]
where ${\rm E}$ is a momentum canonically conjugate to $A_1$.

On the circle boundary conditions for the fields must be specified.
We impose the following ones
\begin{eqnarray}
{A_1} (- \frac{\rm L}{2}) & = & {A_1} (\frac{\rm L}{2}) \nonumber \\
{\psi_{\rm R}} (- \frac{\rm L}{2}) & = & {\psi_{\rm R}} (\frac{\rm L}{2}).
\label{eq: tri}
\end{eqnarray}

The Lagrangian density ~\ref{eq: odin} and the Hamiltonian density
{}~\ref{eq: ddva} are invariant under local time-independent transformations
\begin{eqnarray*}
A_1 & \rightarrow & A_1 + {\partial}_{1} \lambda,\\
\psi_{\rm R} & \rightarrow & \exp\{ie \lambda\} \psi_{\rm R},
\end{eqnarray*}
generated by
\[
{\rm G}  = \partial_{1} {\rm E} + e j_{\rm R},
\]
as well as under global gauge transformations of the right-handed
Dirac field which are generated by
\[ {\rm Q}_{\rm R} = \int_{-{\rm L}/2}^{{\rm L}/2} dx j_{\rm R}(x) ,\]
where $j_{\rm R} = \hbar \psi_{\rm R}^{\star} \psi_{\rm R}$ is the classical
right-handed fermionic current and $\lambda (x)$ is a gauge function.

Due to the gauge invariance, the Hamiltonian density is not unique.
On the constrained submanifold ${\rm G} \approx 0$ of the full phase space,
the Hamiltonian density
\begin{equation}
{\cal H}_{\rm R}^{\star} = {\cal H}_{\rm R} + v_{\rm H} \cdot {\rm G},
\label{eq: cet}
\end{equation}
where $v_{\rm H}$ is an arbitrary Lagrange multiplier depending generally
on field variables and their momenta, reduces to the Hamiltonian density
${\cal H}_{\rm R}$. In this sense, there is no difference between
${\cal H}_{\rm R}$ and ${\cal H}_{\rm R}^{\star}$ , and so both
Hamiltonian densities are physically equivalent to each other.

The gauge transformations which respect the boundary conditions
{}~\ref{eq: tri} must be of the form
\begin{equation}
\lambda (\frac{\rm L}{2})=\lambda (- \frac{\rm L}{2}) + {\frac{2\pi}{e}}n,
\hspace{5 mm}
{\rm n} \in \cal Z.
\label{eq: pet}
\end{equation}
We see that the gauge transformations under consideration are divided
into topological classes characterized by the integer $n$. If
$ \lambda (\frac{\rm L}{2}) = \lambda (- \frac{\rm L}{2})$, then the gauge
transformation is topologically trivial and belongs to the $n=0$
class. If $n \neq 0$ it is nontrivial and has winding number $n$.

Given Eq.~\ref{eq: pet}, the nonintegrable phase
\[
\Gamma(A)=\exp\{\frac{ie}{2\pi}\int_{-{\rm L}/2}^{{\rm L}/2}dx {A_1}(x,t) \}
\]
is a unique gauge-invariant quantity that can be constructed from the
gauge field \cite{mant85,raje88,hetr88,sarad88}.
By a topologically trivial transformation we can make
$A_1$ independent of $x$,
\[ {A_1}(x,t) = b(t) ,\]
i.e. obeying the Coulomb gauge ${\partial_1}A_1=0$, then
\[ \Gamma (A) = \exp\{i \frac{e{\rm L}}{2\pi} b(t)\} .\]
In contrast to $\Gamma (A)$ , the line integral
\[ b(t)={\frac{1}{{\rm L}}}\int_{-{\rm L}/2}^{{\rm L}/2} dx {A_1}(x,t) \]
is invariant only under the topologically trivial gauge transformations.
The gauge transformations from the $n$th topological class shift $b$
by $\frac{2\pi}{e{\rm L}}n$. By a non-trivial gauge transformation of the
form $g_n=\exp\{i\frac{2\pi}{\rm L}nx\}$, we can then bring $b$ into the
interval $[0 , \frac{2\pi}{e{\rm L}}]$ . The configurations $b=0$ and
$b=\frac{2\pi}{e{\rm L}}$ are gauge equivalent, since they are connected
by the gauge transformation from the first topological class. The
gauge-field configuration is therefore a circle with length
$\frac{2\pi}{e{\rm L}}$.

\subsection{QUANTIZATION AND ANOMALY}

The eigenfunctions and the eigenvalues of the first quantized
fermionic Hamiltonian are
\[ d \langle x|n;{\rm R} \rangle = \varepsilon_{n,{\rm R}}
\langle x|n;{\rm R} \rangle ,\]
where
\[
\langle x|n;{\rm R} \rangle = \frac{1}{\sqrt {\rm L}}
\exp\{ie\int_{-{\rm L}/2}^{x} dz{A_1}(z)+
i\varepsilon_{n,{\rm R}} \cdot x\},
\]
\[
\varepsilon_{n,{\rm R}} = \frac{2\pi}{\rm L}
(n - \frac{eb{\rm L}}{2\pi}).
\]
We see that the energy spectrum depends on $b$. For $\frac{eb{\rm L}}{2\pi}
={\rm integer}$, the spectrum contains the zero energy level. As $b$
increases from $0$ to $\frac{2\pi}{e{\rm L}}$, the energies of $\varepsilon
_{n,\rm R}$ decrease by $\frac{2\pi}{\rm L}$. Some of energy levels change
sign. However, the spectrum at the configurations $b=0$ and
$b=\frac{2\pi}{e{\rm L}}$ is the same, namely, the integers, as it must be
since these gauge-field configurations are gauge-equivalent. In what
follows, we will use separately the integer and fractional parts of
$\frac{eb{\rm L}}{2\pi}$, denoting them as $[\frac{eb{\rm L}}{2\pi}]$ and
$\{ \frac{eb{\rm L}}{2\pi} \}$ correspondingly.

Now we introduce the second quantized right-handed Dirac field.
For the moment, we will assume that $d$ does not have zero eigenvalue.
At time $t=0$, in terms of the eigenfunctions of the first quantized
fermionic Hamiltonian the second quantized ($\zeta$--function
regulated) field has the expansion \cite{niese86} :
\begin{equation}
\psi_{\rm R}^s (x) = \sum_{n \in \cal Z} a_n \langle x|n;{\rm R} \rangle
|\lambda \varepsilon_{n,\rm R}|^{-s/2}.
\label{eq: shest}
\end{equation}
Here $\lambda$ is an arbitrary constant with dimension of length
which is necessary to make $\lambda \varepsilon_{n,\rm R}$ dimensionless,
while $a_n, a_n^{\dagger}$ are right-handed fermionic creation and
annihilation operators which fulfil the commutation relations
\[  [a_n , a_m^{\dagger}]_{+} = \delta_{m,n} .\]
For $\psi_{\rm R}^{s} (x)$, the equal time anticommutator is
\begin{equation}
[\psi_{\rm R}^{s}(x) , \psi_{\rm R}^{\dagger s}(y)]_{+} = \zeta (s,x,y),
\label{eq: sem}
\end{equation}
with all other anticommutators vanishing, where
\[ \zeta(s,x,y) \equiv \sum_{n \in \cal Z} \langle x|n;{\rm R} \rangle
\langle n;{\rm R}|y \rangle |\lambda \varepsilon_{n,\rm R}|^{-s},\]
$s$ being large and positive. In the limit, when the regulator
is removed, i.e. $s=0$, $\zeta(s=0,x,y) = \delta(x-y)$ and
Eq.~\ref{eq: sem} takes the standard form.

The vacuum state of the second quantized fermionic Hamiltonian
is defined such that all negative energy levels are filled:
\begin{eqnarray}
a_n |{\rm vac};A \rangle =0 & {\rm for} & n>[\frac{eb{\rm L}}{2\pi}],
\nonumber \\
a_n^{\dagger} |{\rm vac};A \rangle =0 & {\rm for} & n \leq
[\frac{eb{\rm L}}{2\pi}].
\label{eq: vosem}
\end{eqnarray}
i.e. the levels with energy lower than (and equal to) the energy
of the level $n=[\frac{eb{\rm L}}{2\pi}]$ are filled and the others
are empty. Excited states are constructed by operating creation
operators on the Fock vacuum.

In the $\zeta$--function regularization scheme, we define the
action of the functional derivative on first quantized fermionic
kets and bras by
\begin{eqnarray*}
\frac{\delta}{\delta {A_1}(x)} |n;{\rm R} \rangle & = & \lim_{s \to 0}
\sum_{m \in \cal Z} |m;{\rm R} \rangle \langle m;{\rm R}| \frac{\delta}
{\delta {A_1}(x)} |n;{\rm R} \rangle |\lambda \varepsilon_{m,\rm R}|^{-s/2},\\
\langle n;{\rm R}| \frac{\stackrel{\leftarrow}{\delta}}{\delta {A_1}(x)}
& = & \lim_{s \rightarrow 0} \sum_{m \in \cal Z} \langle n;{\rm R}|
\frac{\stackrel{\leftarrow}{\delta}}{\delta {A_1}(x)} |m;{\rm R} \rangle
\langle m;{\rm R}| |\lambda \varepsilon_{m,\rm R}|^{-s/2}.
\end{eqnarray*}
 From ~\ref{eq: shest} we get the action of $\frac{\delta}
{\delta {A_1}(x)}$ on the operators $a_n$, $a_n^{\dagger}$
in the form
\begin{eqnarray}
\frac{\delta}{\delta {A_1}(x)} a_n & = & - \lim_{s \to 0}
\sum_{m \in \cal Z} \langle n;{\rm R}| \frac{\delta}{\delta {A_1}(x)}
|m;{\rm R}\rangle a_m |\lambda \varepsilon_{m,\rm R}|^{-s/2},\nonumber \\
\frac{\delta}{\delta {A_1}(x)} a_n^{\dagger} & = & \lim_{s \to 0}
\sum_{m \in \cal Z} \langle m;{\rm R}| \frac{\delta}{\delta {A_1}(x)}
|n;{\rm R} \rangle a_m^{\dagger} |\lambda \varepsilon_{m,\rm R}|^{-s/2}.
\label{eq: devet}
\end{eqnarray}
Next we define the quantum right-handed fermionic current and
fermionic part of the second-quantized Hamiltonian as
\begin{equation}
\hat{j}_{\rm R}^s(x) = \frac{1}{2} \hbar [\psi_{\rm R}^{\dagger s}(x),
\psi_{\rm R}^{s}(x)]_{-}
\label{eq: deset}
\end{equation}
and
\begin{equation}
\hat{\rm H}_{\rm F}^s = \int dx {\cal H}_{\rm F}^s =
\frac{1}{2} \hbar \int dx (\psi_{\rm R}^{\dagger s} d \psi_{\rm R}^s
- \psi_{\rm R}^s d^{\star} \psi_{\rm R}^{\dagger s}).
\label{eq: odinodin}
\end{equation}
Substituting ~\ref{eq: shest} into ~\ref{eq: deset} and ~\ref{eq: odinodin},
we get
\begin{eqnarray*}
\hat{j}_{\rm R}^s(x) & = & \hbar \sum_{n \in \cal Z} \frac{1}{\rm L}
\exp\{i \frac{2 \pi}{\rm L} nx\} \rho_{s}(n),\\
\rho_{s}(n) & \equiv & \sum_{k \in \cal Z} \frac{1}{2} [a_k^{\dagger},
a_{k+n}]_{-} \cdot |\lambda \varepsilon_{k, \rm R}|^{-s/2}
|\lambda \varepsilon_{k+n,\rm R}|^{-s/2}
\end{eqnarray*}
and
\[
\hat{\rm H}_{\rm F}^s = \hbar \sum_{n \in \cal Z}  \frac{1}{\rm L}
\exp\{i\frac{2\pi}{\rm L}nx\} {\cal H}_{\rm F}^s(n),
\]
\begin{equation}
{\cal H}_{\rm F}^s(n) \equiv {\cal H}_0^s(n) - eb{\rho}_{s}(n),
\label{eq: dopdva}
\end{equation}
\[
{\cal H}_0^s(n) \equiv \frac{\pi}{\rm L} \sum_{k \in \cal Z}
(2k+p) \cdot \frac{1}{2} [a_k^{\dagger}, a_{k+p}]_{-} \cdot
|\lambda \varepsilon_{k, \rm R}|^{-s/2}
|\lambda \varepsilon_{k+p , \rm R}|^{-s/2} .
\]
The charge corresponding to the current $\hat{j}_{\rm R}^s(x)$ is
\begin{equation}
\hat{\rm Q}_{\rm R}^s = \int_{-{\rm L}/2}^{{\rm L}/2} dx
\hat{j}_{\rm R}^s(x) = \hbar \rho_{s}(0).
\label{eq: odindva}
\end{equation}
With Eq.~\ref{eq: vosem}, we have for the vacuum expectation values:
\begin{eqnarray*}
\langle {\rm vac},A| \hat{j}_{\rm R}(x)  |{\rm vac},A \rangle & = &
-\frac{1}{2} \hbar \eta_{\rm R},\\
\langle {\rm vac},A| \hat{\rm H}_{\rm F} |{\rm vac},A \rangle & = &
- \frac{1}{2} \hbar \xi_{\rm R},
\end{eqnarray*}
where
\begin{eqnarray}
\eta_{\rm R} & \equiv & \lim_{s \to 0} \frac{1}{\rm L}
\sum_{k \in \cal Z} {\rm sign}(\varepsilon_{k,\rm R})
|\lambda \varepsilon_{k,\rm R}|^{-s},\nonumber \\
\xi_{\rm R} & \equiv & \lim_{s \to 0} \frac{1}{\lambda}
\sum_{k \in \cal Z} |\lambda \varepsilon_{k,\rm R}|^{-s+1}.
\label{eq: odintri}
\end{eqnarray}
The operators ~\ref{eq: deset}, ~\ref{eq: odinodin} and
{}~\ref{eq: odindva} can be therefore written as
\begin{eqnarray}
\hat{j}_{\rm R}(x) & = & :\hat{j}_{\rm R}(x):  -  {\frac{1}{2}}
\hbar \eta_{\rm R},\nonumber \\
\hat{\rm Q}_{\rm R} & = & \hbar :\rho (0):  -  \frac{\rm L}{2}
\hbar \eta_{\rm R}, \\
\hat{\rm H}_{\rm F} & = & \hat{\rm H}_0  - eb \hbar :\rho(0): -
\frac{1}{2} \hbar \xi_{\rm R},   \nonumber
\label{eq: buran}
\end{eqnarray}
where double dots indicate normal ordering with respect to
$|{\rm vac},A \rangle$ and
\begin{eqnarray*}
\hat{\rm H}_0 & = & \hbar \frac{2 \pi}{\rm L} \lim_{s \to 0}
\{ \sum_{k >[\frac{eb{\rm L}}{2 \pi}]} k a_k^{\dagger} a_k
|\lambda \varepsilon_{k,\rm R}|^{-s}  -  \sum_{k \leq [\frac{eb{\rm L}}
{2 \pi}]} k a_k a_k^{\dagger} |\lambda \varepsilon_{k,\rm R}|^{-s} \},\\
:\rho (0): & \equiv & \frac{1}{\hbar} \hat{\rm Q}_{\rm R,\rm N} =
\lim_{s \to 0} \{ \sum_{k >[\frac{eb{\rm L}}{2 \pi}]} a_k^{\dagger}
a_k |\lambda \varepsilon_{k,\rm R}|^{-s}  -  \sum_{k \leq
[\frac{eb{\rm L}}{2 \pi}]} a_k a_k^{\dagger} |\lambda \varepsilon_{k,\rm R}|
^{-s} \}.
\end{eqnarray*}
Taking the sums in ~\ref{eq: odintri}, we get
\begin{eqnarray*}
\eta_{\rm R} & = & \frac{2}{\rm L} ( \{ \frac{eb{\rm L}}{2 \pi} \}
- \frac{1}{2} ), \\
\xi_{\rm R} & = & - \frac{2 \pi}{\rm L}
( ( \{ \frac{eb{\rm L}}{2 \pi} \}
- \frac{1}{2} )^2 - \frac{1}{12} ).
\end{eqnarray*}
Both operators $:\hat{j}_{\rm R}(x):$ and $:\hat{\rm H}_{\rm F}:$ are
well defined when acting on finitely excited states which have only a
finite number of excitations relative to the Fock vacuum.

To construct the quantized electromagnetic Hamiltonian, we first
introduce the Fourier expansion for the gauge field
\begin{equation}
{A_1}(x) = b + \sum_{\stackrel {p \in \cal Z}{p \neq 0}}
e^{i \frac{2 \pi}{\rm L} px} \alpha_p.
\label{eq: odinshest}
\end{equation}
Since ${A_1}(x)$ is a real function, $\alpha_p$ satisfies
\[ \alpha_p = \alpha_{-p}^{\star}. \]
The Fourier expansion for the canonical momentum conjugate to ${A_1}(x)$
is then
\[
\hat{\rm E}(x) = \frac{1}{\rm L} \hat{\pi}_b - \frac{i}{\rm L} \hbar
\sum_{\stackrel {p \in \cal Z} {p \neq 0}}
e^{-i \frac{2\pi}{\rm L}px} \frac{d}{d{\alpha_p}} ,
\]
where $\hat{\pi}_b \equiv -i \hbar \frac{d}{db}$ .
The electromagnetic part of the Hamiltonian density is
\[
\hat{\cal H}_{\rm EM}(x) = \hbar \sum_{p \in \cal Z} \frac{1}{\rm L}
\exp\{i\frac{2\pi}{\rm L}px\} \cdot {\cal H}_{\rm EM}(p),
\]
where
\begin{equation}
{\cal H}_{\rm EM}(p) \equiv - \frac{1}{\rm L} \hbar
\frac{d}{d{\alpha}_{-p}} \frac{d}{db} - \frac{1}{2\rm L} \hbar
\sum_{\stackrel{q \in \cal Z}{q \neq (0;p)}}
\frac{d}{d{\alpha}_{-p+q}} \frac{d}{d{\alpha}_{-q}}
\hspace{1 cm} (p \neq 0) ,
\label{eq: dopodin}
\end{equation}
so the corresponding quantum Hamiltonian becomes
\[
\hat{\rm H}_{\rm EM} = \hbar {\cal H}_{\rm EM}(p=0) = \frac{1}{2\rm L}
\hat{\pi}_b^2 - \frac{1}{\rm L} {\hbar}^2 \sum_{p >0}
\frac{d}{d{\alpha_p}} \frac{d}{d{\alpha_{-p}}}.
\]
The total quantum Hamiltonian is
\[
\hat{\rm H}_{\rm R} = \hat{\rm H}_0 + \hat{\rm H}_{\rm EM}
-eb\hat{\rm Q}_{\rm R,N} - \frac{1}{2} \hbar \xi_{\rm R}.
\]

If we multiply two operators that are finite linear combinations of
the fermionic creation and annihilation operators, the $\zeta$--function
regulated operator product agrees with the naive product. However, if
the operators involve infinite summations their naive product is not
generally well defined. We then define the operator product by
mutiplying the regulated operators with $s$ large and positive and
analytically continue the result to $s=0$. In this way we obtain the
following relations (see Appendix )
\begin{equation}
[\rho(m) , \rho(n)]_{-} = m \delta_{m,-n},
\label{eq: odinsem}
\end{equation}
\begin{equation}
[\hat{\rm H}_0 , \rho(m)]_{-} = - \hbar \frac{2 \pi}{\rm L} m\rho(m),
\label{eq: odinvosem}
\end{equation}
and
\begin{eqnarray}
\frac{d}{db}  \rho(m) & = & 0,\nonumber \\
\frac{d}{d{\alpha_{\pm p}}}  \rho(m) & = & \mp \frac{e\rm L}{2 \pi}
\delta_{p, \pm m} ,    \hspace{1 cm}  (p > 0).
\label{eq: dvanol}
\end{eqnarray}

The quantum Gauss operator is
\[
\hat{\rm G} = \hat{\rm G}_0 + \frac{2 \pi}{{\rm L}^2}
\sum_{p>0} \{ \hat{\rm G}_{+}(p) e^{i\frac{2 \pi}{\rm L} px} -
\hat{\rm G}_{-}(p) e^{-i\frac{2\pi}{\rm L} px} \},
\]
where
\begin{eqnarray*}
\hat{\rm G}_0 & \equiv & \frac{e}{\rm L} \hbar \rho (0),\\
\hat{\rm G}_{\pm}(p) & \equiv & \hbar (p\frac{d}{d{\alpha}_{\mp p}} \pm
\frac{e\rm L}{2\pi} \rho(\pm p)).
\end{eqnarray*}
Using ~\ref{eq: odinsem} and ~\ref{eq: dvanol}, we easily get
that $\rho (\pm p)$ are gauge-invariant:
\[
[\hat{\rm G}_{+}(p), \rho(\pm q) ]_{-} = 0,
\]
\[
[\hat{\rm G}_{-}(p), \rho(\pm q) ]_{-} = 0,
\]
$(p>0, q>0)$. The operators $\hat{\rm G}_{\pm}(p)$ don't commute
with themselves,
\begin{equation}
[\hat{\rm G}_{+}(p) , \hat{\rm G}_{-}(q)]_{-} =
{\hbar}^2 \frac{e^2{\rm L}^2}{4{\pi}^2} p \delta_{p,q}
\label{eq: dvaodin}
\end{equation}
as well as with the Hamiltonian
\begin{equation}
[\hat{\rm H}_{\rm R} , \hat{\rm G}_{\pm}(p)]_{-} =
\pm {\hbar}^3 \frac{e^2{\rm L}}{4{\pi}^2}
\frac{d}{d{\alpha_{\mp p}}}  .
\label{eq: dvadva}
\end{equation}
The commutation relations ~\ref{eq: dvaodin} and ~\ref{eq: dvadva}
reflect an anomalous behaviour of the CSM.

\newpage
\section{QUANTUM CONSTRAINTS}
\label{sec: const}

\subsection{QUANTUM SYMMETRY}

In non-anomalous gauge theory, Gauss law is considered to be valid
for physical states only. This identifies physical states as those
which are gauge-invariant. The problem with the anomalous behaviour
of the CSM, in terms of states in Hilbert space, is now apparent
from Eqs.~\ref{eq: dvaodin} -- ~\ref{eq: dvadva} : we cannot require
that states be annihilated by the Gauss law generators $\hat{\rm G}_
{\pm}(p)$.

Let us represent the action of the topologically trivial gauge
transformations by the operator
\begin{equation}
{\rm U}_{0}(\tau) = \exp\{\frac{i}{\hbar} \hat{\rm G}_{0}
{\tau}_0 + \frac{i}{\hbar} \sum_{p>0}
(\hat{\rm G}_{+} \tau_{+} + \hat{\rm G}_{-} \tau_{-}) \}
\label{eq: dvatri}
\end{equation}
with $\tau_0$ , ${\tau}_{\pm}(p)$ smooth, then
\begin{eqnarray*}
{\rm U}_0^{-1}(\tau) {\alpha}_{\pm p} {\rm U}_0(\tau) & = &
{\alpha}_{\pm}  - i p {\tau}_{\mp}(p),\\
{\rm U}_0^{-1}(\tau) \frac{d}{d {\alpha}_{\pm p}} {\rm U}_0(\tau) & = &
\frac{d}{d {\alpha}_{\pm p}} + i(\frac{e\rm L}{2\pi})^2 {\tau}_{\pm}(p).
\end{eqnarray*}
We find from Eq.~\ref{eq: dvaodin} that
\begin{equation}
{\rm U}_0({\tau}^{(1)}) {\rm U}_0({\tau}^{(2)}) =
\exp\{ 2\pi i {\omega}_{2}({\tau}^{(1)}, {\tau}^{(2)})\}
{\rm U}_{0}( {\tau}^{(1)} + {\tau}^{(2)} ) ,
\label{eq: dvacet}
\end{equation}
where
\[
{\omega}_{2}({\tau}^{(1)} , {\tau}^{(2)}) \equiv
- \frac{i}{4\pi} (\frac{e\rm L}{2\pi})^2 \sum_{p>0} p
({\tau }_{-}^{(1)} {\tau }_{+}^{(2)} - {\tau}_{+}^{(1)} {\tau}_{-}^{(2)})
\]
is a two-cocycle of the gauge group algebra. We are thus dealing
with a projective representation.

The two-cocycle ${\omega}_{2}( {\tau}^{(1)} , {\tau}^{(2)} )$
is trivial, since it can be removed  from ~\ref{eq: dvacet} by
a simple redefinition of ${\rm U}_{0}(\tau)$. Indeed, the modified
operators
\begin{equation}
\tilde{\rm U}_0(\tau) = \exp\{i 2\pi {\alpha}_{1}(\gamma ; \tau)\}
\cdot {\rm U}_{0}(\tau),
\label{eq: dvapet}
\end{equation}
where
\[
{\alpha}_{1}(\gamma, \tau) \equiv - \frac{1}{4 \pi}
(\frac{e\rm L}{2\pi})^2
\sum_{p>0}( {\alpha}_{-p}{\tau}_{-} - {\alpha}_{p} {\tau}_{+} )
\]
is a one-cocycle, satisfy the ordinary composition law
\[
\tilde{\rm U}_0({\tau}^{(1)}) \tilde{\rm U}_0({\tau}^{(2)}) =
\tilde{\rm U}_0({\tau}^{(1)} + {\tau}^{(2)}),
\]
i.e. the action of the topologically trivial gauge transformations
represented by ~\ref{eq: dvapet} is unitary.

The modified Gauss law generators corresponding to ~\ref{eq: dvapet}
are
\begin{equation}
\hat{\tilde{\rm G}}_{\pm}(p) = \hat{\rm G}_{\pm}(p)  \pm
\hbar \frac{{e^2}{\rm L}^2}{8{\pi}^2} {\alpha}_{\pm p}.
\label{eq: dvashest}
\end{equation}
The generators $\hat{\tilde{\rm G}}_{\pm}(p)$ commute:
\[
[\hat{\tilde{\rm G}}_{+}(p) , \hat{\tilde{\rm G}}_{-}(q)]_{-}=0.
\]
This means that Gauss law can be maintained at the quantum level. We
define physical  states as those which are annihilated by
$\hat{\tilde{\rm G}}_{\pm}(p)$   {\cite{sarad91}} :
\begin{equation}
\hat{\tilde{\rm G}}_{\pm}(p) |{\rm phys}; A \rangle = 0.
\label{eq: dvasem}
\end{equation}
The zero component ${\hat{\rm G}}_0$ is a quantum generator of the
global gauge transformations of the right-handed fermionic field,
so the other quantum constraint is
\begin{equation}
\hat{\rm Q}_{\rm R , \rm N} |{\rm phys};A \rangle = 0.
\label{eq: dvavosem}
\end{equation}

\subsection{ADIABATIC APPROACH}

Let us show now that we can come to the quantum constraints
{}~\ref{eq: dvasem} and ~\ref{eq: dvavosem} in a different way, using
the adiabatic approximation  \cite{schiff68,berry84}.
In the adiabatic approach, the dynamical variables are divided
into two sets, one which we call fast variables and the other
which we call slow variables. In our case, we treat the fermions
as fast variables and the gauge fields as slow variables.

Let ${\cal A}^1$ be a manifold of all static gauge field
configurations ${A_1}(x)$. On ${\cal A}^1$  a time-dependent
gauge field ${A_1}(x,t)$ corresponds to a path and a periodic gauge
field to a closed loop.

We consider the fermionic part of the second-quantized Hamiltonian
$:\hat{\cal H}_{\rm F}:$ which depends on $t$  through the background
gauge field $A_1$ and so changes very slowly with time. We consider
next the periodic gauge field ${A_1}(x,t) (0 \leq t <T)$ . After a
time $T$ the periodic field ${A_1}(x,t)$ returns to its original
value: ${A_1}(x,0) = {A_1}(x,T)$, so that $:\hat{\rm H}_{\rm F}:(0)=
:\hat{\rm H}_{\rm F}:(T)$ .

At each instant $t$ we define eigenstates for $:\hat{\rm H}_{\rm F}:
(t)$ by
\[
:\hat{\rm H}_{\rm F}:(t) |{\rm F}, A(t) \rangle =
{\varepsilon}_{\rm F}(t) |{\rm F}, A(t) \rangle.
\]
The state $|{\rm F}=0, A(t) \rangle \equiv |{\rm vac}, A(t) \rangle$
is a ground state of $:\hat{\rm H}_{\rm F}:(t)$ :
\[
:\hat{\rm H}_{\rm F}:(t) |{\rm vac}, A(t) \rangle =0.
\]
The Fock states $|{\rm F}, A(t) \rangle $ depend on $t$ only through
their implicit dependence on $A_1$. They are assumed to be
orthonormalized,
\[
\langle {\rm F^{\prime}}, A(t)|{\rm F}, A(t) \rangle =
\delta_{{\rm F},{\rm F^{\prime}}},
\]
and nondegenerate.

The time evolution of the wave function  of our system (fermions
in a background gauge field) is clearly governed by the Schrodinger
equation:
\[
i \hbar \frac{\partial \psi(t)}{\partial t} =
:\hat{\rm H}_{\rm F}:(t) \psi(t) .
\]
For each $t$, this wave function can be expanded in terms of the
"instantaneous" eigenstates $|{\rm F}, A(t) \rangle$ .

Let us choose ${\psi}_{\rm F}(0)=|{\rm F}, A(0) \rangle$, i.e.
the system is initially described by the eigenstate
$|{\rm F},A(0) \rangle$ . According to the adiabatic approximation,
if at $t=0$ our system starts in an stationary state $|{\rm F},A(0)
\rangle $ of $:\hat{\rm H}_{\rm F}:(0)$, then it will remain,
at any other instant of time $t$, in the corresponding eigenstate
$|{\rm F}, A(t) \rangle$ of the instantaneous Hamiltonian
$:\hat{\rm H}_{\rm F}:(t)$. In other words, in the adiabatic
approximation transitions to other eigenstates are neglected.

Thus, at some time $t$ later our system will be described up to
a phase by the same Fock state $|{\rm F}, A(t) \rangle $:
\begin{equation}
\psi_{\rm F}(t) = {\rm C}_{\rm F}(t) \cdot |{\rm F},A(t) \rangle,
\label{eq: dvadevet}
\end{equation}
where ${\rm C}_{\rm F}(t)$ is yet undetermined phase.

To find this phase, we insert ~\ref{eq: dvadevet} into the
Schrodinger equation :
\begin{equation}
\hbar \dot{\rm C}_{\rm F}(t) = -i {\rm C}_{\rm F}(t)
\varepsilon_{\rm F}(t) - \hbar {\rm C}_{\rm F}(t)
\langle {\rm F},A(t)|\frac{\partial}{\partial t}|{\rm F},A(t) \rangle.
\label{eq: trinol}
\end{equation}
Solving ~\ref{eq: trinol}, we get
\[
{\rm C}_{\rm F}(t) = \exp\{- \frac{i}{\hbar} \int_{0}^{t} d{t^{\prime}}
{\varepsilon}_{\rm F}({t^{\prime}}) - \int_{0}^{t} d{t^{\prime}}
\langle {\rm F},A({t^{\prime}})|\frac{\partial}{\partial{t^{\prime}}}|
{\rm F},A({t^{\prime}}) \rangle \}.
\]
For $t=T$, $|{\rm F},A(T) \rangle =|{\rm F},A(0) \rangle$ ( the
instantaneous eigenfunctions are chosen to be periodic in time)
and
\[
{\psi}_{\rm F}(T) = \exp\{i {\gamma}_{\rm F}^{\rm dyn} +
i {\gamma}_{\rm F}^{\rm Berry} \}\cdot {\psi}_{\rm F}(0),
\]
where
\[ {\gamma}_{\rm F}^{\rm dyn} \equiv - \frac{1}{\hbar}
\int_{0}^{T} dt \cdot {\varepsilon}_{\rm F}(t) ,    \]
while
\begin{equation}
{\gamma}_{\rm F}^{\rm Berry} \equiv \int_{0}^{T} dt \int_{-{\rm L}/2}^
{{\rm L}/2} dx \dot{A_1}(x,t) \langle {\rm F},A(t)|i \frac{\delta}
{\delta A_1(x,t)}|{\rm F},A(t) \rangle
\label{eq: triodin}
\end{equation}
is Berry's phase  \cite{berry84}.

If we define the $U(1)$ connection
\begin{equation}
{\cal A}_{\rm F}(x,t) \equiv \langle {\rm F},A(t)|i \frac{\delta}
{\delta A_1(x,t)}|{\rm F},A(t) \rangle,
\label{eq: tridva}
\end{equation}
then
\[
{\gamma}_{\rm F}^{\rm Berry} = \int_{0}^{T} dt \int_{-{\rm L}/2}^
{{\rm L}/2} dx \dot{A}_1(x,t) {\cal A}_{\rm F}(x,t).
\]
We see that upon parallel transport around a closed loop on
${\cal A}^1$ the Fock state $|{\rm F},A(t) \rangle$ acquires an
additional phase which is integrated exponential of ${\cal A}_{\rm F}
(x,t)$. Whereas the dynamical phase ${\gamma}_{\rm F}^{\rm dyn}$
provides information about the duration of the evolution, the
Berry's phase reflects the nontrivial holonomy of the Fock states
on ${\cal A}^1$.

However, a direct computation of the diagonal matrix elements of
$\frac{\delta}{\delta A_1(x,t)}$ in ~\ref{eq: triodin} requires a
globally single-valued basis for the eigenstates $|{\rm F},A(t) \rangle$
which is not available. The connection ~\ref{eq: tridva} can be
defined only locally on ${\cal A}^1$, in regions where $[\frac{eb{\rm L}}
{2 \pi}]$ is fixed. The values of $A_1$ in regions of different
$[\frac{eb{\rm L}}{2 \pi}]$ are connected by topologically nontrivial
gauge transformations. If $[\frac{eb{\rm L}}{2 \pi}]$ changes, then
there is a nontrivial spectral flow , i.e. some of energy levels
of the first quantized fermionic Hamiltonian cross zero and change
sign. This means that the definition of the Fock vacuum of the second
quantized fermionic Hamiltonian changes (see Eq.~\ref{eq: vosem}).
Since the creation and annihilation operators $a^{\dagger}, a$ are
continuous functionals of $A_1(x)$, the definition of all excited
Fock states $|{\rm F},A(t) \rangle$ is also discontinuous. The
connection ${\cal A}_{\rm F}$ is not therefore well-defined globally.
Its global characterization necessiates the usual introduction of
transition functions.

Furthermore, ${\cal A}_{\rm F}$ is not invariant under $A$--dependent
redefinitions of the phases of the Fock states: $|{\rm F},A(t) \rangle
\rightarrow \exp\{-i \chi[A]\} |{\rm F},A(t) \rangle$, and transforms like
a $U(1)$ vector potential
\[
{\cal A}_{\rm F} \rightarrow {\cal A}_{\rm F} +
\frac{\delta \chi[A]}{\delta A_1}.
\]

For these reasons, to calculate ${\gamma}_{\rm F}^{\rm Berry}$ it
is more convenient to compute first the $U(1)$ curvature tensor
\begin{equation}
{\cal F}_{\rm F}(x,y,t) \equiv \frac{\delta}{\delta A_1(x,t)}
{\cal A}_{\rm F}(y,t) - \frac{\delta}{\delta A_1(y,t)}
{\cal A}_{\rm F}(x,t)
\label{eq: tritri}
\end{equation}
and then deduce ${\cal A}_{\rm F}$.

For simplicity, let us compute the vacuum curvature tensor
${\cal F}_{{\rm F}=0}(x,y,t)$. Substituting ~\ref{eq: tridva}
into ~\ref{eq: tritri}, we get
\[
{\cal F}_{{\rm F}=0}(x,y,t) = i \sum_{{\rm F} \neq 0}
\{ \langle {\rm vac}, A(t)|\frac{\delta}{\delta A_1(y,t)}|
{\rm F},A(t) \rangle \langle {\rm F},A(t)|\frac{\delta}
{\delta A_1(x,t)}|{\rm vac},A(t) \rangle
\]
\begin{equation}
- (x \longleftrightarrow y) \},
\label{eq: tricet}
\end{equation}
where the summation is over the complete set of states
$|{\rm F},A(t) \rangle$.

Using the formula
\[
\langle {\rm vac},A(t) | \frac{\delta}{\delta A_1(x,t)}|
{\rm F}, A(t) \rangle =\frac{1}{{\varepsilon}_{\rm F}}
\langle {\rm vac},A(t)| \frac{\delta :\hat{\rm H}_{\rm F}:(t)}
{\delta A_1(x,t)}|{\rm F},A(t) \rangle,
\]
we rewrite ~\ref{eq: tricet} as
\[
{\cal F}_{{\rm F}=0}(x,y,t) = i \sum_{{\rm F} \neq 0}
\frac{1}{{\varepsilon}_{\rm F}^2} \{ \langle {\rm vac},A(t)|
\frac{\delta :\hat{\rm H}_{\rm F}:(t)}{\delta A_1(y,t)}|
{\rm F},A(t) \rangle \cdot \langle {\rm F},A(t)|
\frac{\delta :\hat{\rm H}_{\rm F}:(t)}{\delta A_1(x,t)}
|{\rm vac}, A(t) \rangle
\]
\begin{equation}
- (x \longleftrightarrow y) \}.
\label{eq: tripet}
\end{equation}

Since $\frac{\delta :\hat{\rm H}_{\rm F}:(t)}{\delta A_1}$ is
quadratic in $a^{\dagger}, a$, only excited states of the type
\[
|{\rm F},A(t) \rangle \longleftrightarrow a_m^{\dagger} a_n
|{\rm vac},A(t) \rangle
\hspace{1 cm}
(n \leq [\frac{eb{\rm L}}{2\pi}],
m > [\frac{eb{\rm L}}{2\pi}])
\]
with $\varepsilon_{\rm F}=\frac{2\pi}{\rm L} \hbar (m-n)$
contribute to ~\ref{eq: tripet} which takes then the form
\[
{\cal F}_{{\rm F}=0}(x,y,t) = i \frac{{\rm L}^2}{4 {\pi}^2}
\sum_{m \neq n} \frac{1}{{\hbar}^2(m-n)^2} \{ \langle{\rm vac},A(t)|
\frac{\delta : \hat{\rm H}_{\rm F}:(t)}{\delta A_1(y,t)}
a_m^{\dagger} a_n |{\rm vac},A(t) \rangle  \cdot
\]
\begin{equation}
\langle {\rm vac},A(t)| a_n^{\dagger} a_m
\frac{\delta :\hat{{\rm H}_{\rm F}}:(t)} {\delta A_1(x,t)}
|{\rm vac},A(t) \rangle - (x \longleftrightarrow y) \}.
\label{eq: trishest}
\end{equation}
With $:\hat{\rm H}_{\rm F}:(t)$ given by  $15$,
Eq.~\ref{eq: trishest} is evaluated as
\begin{equation}
{\cal F}_{{\rm F}=0} = \frac{e^2}{2\pi} \sum_{n>0} \frac{1}{n}
\sin(\frac{2\pi}{\rm L} n(x-y)) = \frac{e^2}{4\pi} \epsilon(x-y)
- \frac{e^2}{2\pi \rm L}(x-y).
\label{eq: trisem}
\end{equation}
The corresponding $U(1)$ connection is easily deduced as
\[
{\cal A}_{{\rm F}=0}(x,t) = -\frac{1}{2} \int_{-{\rm L}/2}
^{{\rm L}/2} dy {\cal F}_{{\rm F}=0}(x,y,t) A_1(y,t).
\]
The Berry phase becomes
\[
{\gamma}_{{\rm F}=0}^{\rm Berry} = - \frac{1}{2} \int_{0}^{T} dt
\int_{-{\rm L}/2}^{{\rm L}/2} dx \int_{-{\rm L}/2}^{{\rm L}/2} dy
\dot{A_1}(x,t) {\cal F}_{{\rm F}=0}(x,y,t) A_1(y,t).
\]
We see that in the limit ${\rm L} \to \infty$, when the second term
in ~\ref{eq: trisem} may be neglected, the $U(1)$ curvature tensor
coincides with that obtained in  \cite{niemi86,semen87},
while the Berry phase is
\[
{\gamma}_{{\rm F}=0}^{\rm Berry} = \int_{0}^{T} dt
\int_{- \infty}^{\infty} dx {\cal L}(x,t),
\]
where
\[
{\cal L}(x,t) \equiv - \frac{e^2}{8 {\pi}^2} \int_{- \infty}^{\infty}
dy \dot{A_1}(x,t) \epsilon(x-y) A_1(y,t)
\]
is a non-local part of the effective Lagrange density of the CSM
\cite{sarad93}.

In terms of the Fourier components, the connection ${\cal A}_
{{\rm F}=0}$ is rewritten as
\begin{eqnarray*}
\langle {\rm vac},A(t)|  \frac{d}{db(t)}  | {\rm vac},A(t) \rangle &
= & 0,\\
\langle {\rm vac},A(t)|  \frac{d}{d{\alpha}_{\pm p}(t)}  |
{\rm vac},A(t) \rangle & \equiv & {\cal A}_{\pm}(p,t) =
\pm \frac{e^2{\rm L}^2}{8{\pi}^2} \frac{1}{p} {\alpha}_{\mp p},
\end{eqnarray*}
so the nonvanishing curvature is
\[
{\cal F}_{+ -} \equiv \frac{d}{d{\alpha}_{-p}} {\cal A}_{+} -
\frac{d}{d{\alpha}_{p}} {\cal A}_{-} = \frac{e^2{\rm L}^2}
{4{\pi}^2} \frac{1}{p} .
\]
A parallel transportation of the vacuum $|{\rm vac},A(t) \rangle$
around a closed loop in $({\alpha}_{p}, {\alpha}_{-p})$ --
space $(p>0)$ yields back the same vacuum state multiplied by the phase
\[
{\gamma}_{{\rm F}=0}^{\rm Berry} = \frac{e^2{\rm L}^2}{4{\pi}^2}
\int_{0}^{T} dt \sum_{p>0} \frac{1}{p}
i{\alpha}_{p} \dot{\alpha}_{-p}.
\]
However, this phase associated with the projective representation
of the gauge group is trivial, since it can be removed. If we redefine
the momentum operators as
\begin{equation}
\frac{d}{d{\alpha}_{\pm p}} \longrightarrow
\frac{\tilde{d}}{d{\alpha}_{\pm p}} \equiv \frac{d}{d{\alpha}_{\pm p}}
\mp  \frac{e^2{\rm L}^2}{8{\pi}^2} \frac{1}{p} {\alpha}_{\mp p},
\label{eq: trivosem}
\end{equation}
then the corresponding connection and curvature vanish:
\begin{eqnarray*}
\tilde{{\cal A}}_{\pm} & \equiv & \langle {\rm vac},A(t)|
\frac{\tilde{d}}{d{\alpha}_{\pm p}} |{\rm vac},A(t) \rangle =0,\\
\tilde{\cal F}_{+ -} & = & \frac{\tilde{d}}{d{\alpha}_{-p}}
\tilde{\cal A}_{+} - \frac{\tilde{d}}{d{\alpha}_{p}}
\tilde{\cal A}_{-} =0.
\end{eqnarray*}
The modified momentum operators  are noncommuting:
\[
[\frac{\tilde{d}}{d{\alpha}_{p}} , \frac{\tilde{d}}
{d{\alpha}_{-q}} ]_{-} = \frac{e^2{\rm L}^2}{4{\pi}^2}
\frac{1}{p} {\delta}_{p,q}.
\]
Following ~\ref{eq: trivosem}, we modify the Gauss law generators
as
\[
\hat{\rm G}_{\pm}(p) \longrightarrow \hat{\tilde{\rm G}}_{\pm}(p) =
\hbar(\frac{\tilde{d}}{d{\alpha}_{\mp p}} \pm \frac{e{\rm L}}{2\pi}
\rho(\pm p))
\]
that coincides with ~\ref{eq: dvashest}. The modified Gauss law
generators have vanishing vacuum  expectation values,
\[
\langle {\rm vac},A(t)| \hat{\tilde{\rm G}}_{\pm}(p,t) |
{\rm vac},A(t) \rangle =0.
\]
This justifies the definition ~\ref{eq: dvasem}.

For the zero component $\hat{\rm G}_0$, the vacuum expectation
value
\[
\langle {\rm vac},A(t)| \hat{\rm G}_0 | {\rm vac},A(t) \rangle
= - \hbar \frac{e}{2} {\eta}_{\rm R}
\]
can be also made equal to zero by the redefinition
\[
\hat{\rm G}_0 \longrightarrow \hat{\tilde{\rm G}}_0 +
\hbar \frac{e}{2} {\eta}_{\rm R} = \frac{e}{\rm L} \hbar
:\rho(0):
\]
that leads to ~\ref{eq: dvavosem}.

Thus, both quantum constraints ~\ref{eq: dvasem} and
{}~\ref{eq: dvavosem} can be realized in the framework of the
adiabatic approximation.

\newpage
\section{PHYSICAL QUANTUM CSM}
\label{sec: exoti}
\subsection{CONSTRUCTION OF PHYSICAL HAMILTONIAN}

$1$. From the point of view of Dirac quantization, there are many
physically equivalent classical theories of a system with
first-class constraints. As mentioned, the origin of such an
ambiguity lies in a gauge freedom. For the classical CSM, the
gauge freedom is characterized by an arbitrary $v_{\rm H}(x)$
in ~\ref{eq: cet}. If we use the Fourier expansion for
$v_{\rm H}(x)$, then the general form of the classical Hamiltonian
is rewritten as
\begin{equation}
\tilde{\rm H}_{\rm R}= {\rm H}_{\rm R} +
\sum_{p>0} ( v_{\rm H,+} {\rm G}_{+} + v_{\rm H,-} {\rm G}_{-}) .
\label{eq: tridevet}
\end{equation}
Any Hamiltonian $\tilde{\rm H}_{\rm R}$ with fixed nonzero
$(v_{\rm H,-} , v_{\rm H,+})$ gives rise to the same weak equations
of motion as those deduced from ${\rm H}_{\rm R}$, although the
strong form of these equations may be quite different. The physics is
however described by the weak equations. Different $(v_{\rm H,-},
v_{\rm H,+})$ lead to different mathematical descriptions of the
same physical situation.

To construct the quantum theory of any system with first-class
constraints, we usually quantize one of the corresponding classical
theories. All the possible quantum theories constructed in this way
are believed to be equivalent to each other.

In the case, when gauge degrees of freedom are anomalous, the
situation is different: the physical equivalence of quantum
Hamiltonians is lost. For the CSM, the quantum Hamiltonian
$\hat{\tilde{\rm H}}_{\rm R}$ does not reduce to $\hat{\rm H}_{\rm R}$
on the physical states:
\[
\hat{\tilde{\rm H}}_{\rm R} |{\rm phys};A \rangle \neq
\hat{\rm H}_{\rm R} |{\rm phys};A \rangle .
\]
The quantum theory consistently describing the dynamics of the CSM
should be definitely compatible with ~\ref{eq: dvasem}. The
corresponding quantum Hamiltonian is then defined by the conditions
\begin{equation}
[\hat{\tilde{\rm H}}_{\rm R} , \hat{\tilde{\rm G}}_{\pm}(p)]_{-} =0
\hspace{1 cm} (p>0)
\label{eq: cetnol}
\end{equation}
which specify that $\hat{\tilde{\rm H}}_{\rm R}$ must be invariant
under the modified topologically trivial gauge transformations
generated by $\hat{\tilde{\rm G}}_{\pm}(p)$.

The conditions ~\ref{eq: cetnol} can be considered as a system
of equations for the Lagrange multipliers $\hat{v}_{\rm H,\pm}$
which become operators at the quantum level.
These equations are
\[
\sum_{q>0} \{ \hat{\rm G}_{+}(q) [\hat{v}_{\rm H,+}(q) ,
\hat{\tilde{\rm G}}_{\pm}(p)]_{-} + \hat{\rm G}_{-}(q)[\hat{v}_{\rm H,-}(q),
\hat{\tilde{\rm G}}_{\pm}(p)]_{-} \} \mp {\hbar}^2
\frac{e^2{\rm L}^2}{8{\pi}^2} p \hat{v}_{\rm H,\mp}(p)
\]
\[
= \mp {\hbar}^3 \frac{e^2{\rm L}}{8{\pi}^2}
\frac{d}{d{\alpha}_{\mp p}}
\]
and fix $\hat{v}_{\rm H, \pm}(p)$ in the form
\[
\hat{v}_{\rm H,\pm}(p) = \pm \frac{e}{2\pi} \hbar \frac{1}{p^2}
(\rho(\mp p) + \frac{e\rm L}{8\pi} {\alpha}_{\mp p} ).
\]
Substituting $\hat{v}_{\rm H,\pm}(p)$ into the quantum counterpart
of ~\ref{eq: tridevet}, on the physical states $|{\rm phys};A \rangle$
we get
\begin{equation}
\frac{1}{2} \sum_{p>0} \{ [\hat{v}_{\rm H,+}(p), \hat{\rm G}_{+}(p)]_{+}
+ [\hat{v}_{\rm H,-}(p) , \hat{\rm G}_{-}(p)]_{+} \} = \nonumber \\
\frac{1}{{\rm L}^2} {\hbar}^2 \sum_{p>0} (\frac{d}{d{\alpha}_{p}}
\frac{d}{d{\alpha}_{-p}} - \frac{1}{2} [\frac{\tilde{d}}{d{\alpha}_{p}},
\frac{\tilde{d}}{d{\alpha}_{-p}}]_{+} ),
\label{eq: cetodin}
\end{equation}
i.e. the last term in the right-hand side of ~\ref{eq: tridevet}
contributes only to the electromagnetic part of the Hamiltonian,
changing $\frac{d}{d{\alpha}_{\pm}}$ by $\frac{\tilde{d}}{d{\alpha}_{\pm}}$:
\[
\hat{\rm H}_{\rm EM} \rightarrow \frac{1}{2\rm L} \hat{\pi}_{b}^{2} -
\frac{1}{2\rm L} {\hbar}^2 \sum_{p>0}[\frac{\tilde{d}}{d{\alpha}_{p}},
\frac{\tilde{d}}{d{\alpha}_{-p}}]_{+} .
\]

$2$. The topologically nontrivial gauge transformations change the integer
part of $\frac{eb\rm L}{2\pi}$ :
\begin{eqnarray*}
[\frac{eb\rm L}{2\pi}] & \rightarrow & [\frac{eb\rm L}{2\pi}] + n,\\
\{ \frac{eb\rm L}{2\pi} \} & \rightarrow & \{ \frac{eb\rm L}{2\pi} \},\\
\hat{\psi}_{\rm R} & \rightarrow & \exp\{i\frac{2{\pi}n}{\rm L} x\}
\hat{\psi}_{\rm R}.
\end{eqnarray*}
The action of the topologically nontrivial gauge transformations on
the states can be represented by the operators
\begin{equation}
{\rm U}_n = \exp\{-\frac{i}{\hbar} n \cdot \hat{\rm T}_{b}\}\cdot {\rm U}_0
\label{eq: cetdva}
\end{equation}
where
\[
\hat{\rm T}_{b} \equiv \hat{\pi}_{[\frac{eb\rm L}{2\pi}]} -
\frac{2\pi}{\rm L} \int_{-{\rm L}/2}^{{\rm L}/2} dx x \cdot
\hat{j}_{\rm R}(x) \equiv
-i \hbar \frac{d}{d[\frac{eb\rm L}{2\pi}]} + i \hbar
\sum_{\stackrel{p \in \cal Z}{p \neq 0}} \frac{(-1)^p}{2p} \rho(p)
\]
and ${\rm U}_0$ is given by ~\ref{eq: dvatri}.

To identify the gauge transformation as belonging to the $n$th
topological class we use the index $n$ in ~\ref{eq: cetdva}. The
case $n=0$ corresponds to the topologically trivial gauge
transformations.

The Fourier components of the fermionic current are transformed as
\[
\rho(\pm p)  \rightarrow  \rho(\pm p) - (-1)^p \cdot n ,
\hspace{1 cm} (p>0) .
\]

The composition law ~\ref{eq: dvacet} is valid for the topologically
nontrivial gauge transformations, too. The modified topologically
nontrivial gauge transformation operators are
\[
\tilde{\rm U}_n = \exp\{- \frac{i}{\hbar} n \cdot \hat{\rm T}_b\}
\cdot \tilde{\rm U}_0.
\]
On the physical states
\[
\tilde{\rm U}_n |{\rm phys};A \rangle =
(\exp\{-\frac{i}{\hbar}\hat{\rm T}_{b} \})^n |{\rm phys};A \rangle.
\]

Let $|{\rm phys};A;n \rangle$ be a physical state in which the
integer part of $\frac{eb\rm L}{2\pi}$ is equal $n$. Then in the
state $\exp\{-\frac{i}{\hbar}\hat{\rm T}_b\} |{\rm phys};A;n \rangle=
|{\rm phys};A;n+1 \rangle$ the integer part of $\frac{eb\rm L}{2\pi}$
is equal $n+1$, i.e. the topologically nontrivial gauge transformation
operator $\exp\{-\frac{i}{\hbar}\hat{\rm T}_b\}$ increases
$[\frac{eb\rm L}{2\pi}]$ by one. The operator $\exp\{\frac{i}{\hbar}
\hat{\rm T}_b\}$ decreases $[\frac{eb\rm L}{2\pi}]$ by one.

The vacuum state $|{\rm vac};A;n \rangle$ is defined as follows
\begin{eqnarray}
a_m  |{\rm vac};A;n \rangle = 0 & {\rm for} & m \geq n+1 ,\nonumber \\
a_m^{\dagger}  |{\rm vac};A;n \rangle = 0 & {\rm for} & m<n+1,
\label{eq: cettri}
\end{eqnarray}
the levels with energy lower than ${\varepsilon}_{\rm R,n+1}$
being filled and the others being empty. While the vacuum
{}~\ref{eq: vosem} is defined such that it is always the lowest
energy state at any configuration of the gauge field, the vacuum
{}~\ref{eq: cettri} is the lowest energy state only when the global
gauge field degree of freedom $b$ satisfies the condition
$n \leq \frac{eb\rm L}{2\pi} \leq n+1 $, i.e. $[\frac{eb\rm L}
{2\pi}] = n $ .

Among all states $|{\rm phys};A \rangle$ one may identify the
eigenstates of the operators of the physical variables. The action
of the topologically nontrivial gauge transformations on such
states may, generally speaking, change only the phase of these
states by a C--number, since with any gauge transformations both
topologically trivial and nontrivial, the operators of the physical
variables and the observables cannot be changed. Using
$|{\rm phys}; \theta \rangle$ to designate these physical states,
we have
\begin{equation}
\exp\{\mp \frac{i}{\hbar} \hat{\rm T}_b\}  |{\rm phys}; \theta \rangle
= e^{\pm i\theta}  |{\rm phys}; \theta \rangle,
\label{eq: cetcet}
\end{equation}
The states $|{\rm phys};\theta \rangle$ obeying ~\ref{eq: cetcet}
are easily constructed in the form
\begin{equation}
|{\rm phys};\theta \rangle = \sum_{n \in \cal Z} e^{-in\theta}
(\exp\{-\frac{i}{\hbar} \hat{\rm T}_b\})^n |{\rm phys};A \rangle
\label{eq: cetpet}
\end{equation}
(so called $\theta$--states \cite{jack76,callan76}),
where $|{\rm phys};A \rangle$ is an arbitrary physical state from
{}~\ref{eq: dvasem}.

In one dimension the $\theta$--parameter is related to a constant
background electric field . To show this, we introduce states which
are invariant even against the topologically nontrivial gauge
transformations. Recalling that $[\frac{eb\rm L}{2\pi}]$ is shifted
by $n$ under a gauge transformation from the $n$th topological class,
we easily construct such states as
\begin{equation}
|{\rm phys}\rangle \equiv \exp\{i[\frac{eb\rm L}{2\pi}] \theta\}
|{\rm phys};\theta \rangle.
\label{eq: cetshest}
\end{equation}
The states $|{\rm phys} \rangle$ continue to be annihilated by
$\hat{\tilde{\rm G}}_{\pm}(p)$, and are also invariant under
the topologically nontrivial gauge transformations, so we can
require that
\begin{equation}
\hat{\rm T}_b |{\rm phys} \rangle = 0.
\label{eq: cetsem}
\end{equation}
On the states  ~\ref{eq: cetshest} the electromagnetic part of the
Hamiltonian takes the form
\[
\hat{\rm H}_{\rm EM} \rightarrow \frac{1}{2\rm L}
(\hat{\pi}_b + \hbar \frac{\rm L}{2} {\cal E}_{\theta})^2 -
\frac{1}{2\rm L} {\hbar}^2 \sum_{p>0}
[\frac{\tilde{d}}{d{\alpha}_{p}},\frac{\tilde{d}}{d{\alpha}_{-p}}]_{+},
\]
i.e. the momentum $\hat{\pi}_b$ is supplemented by the electric field
strength ${\cal E}_{\theta} \equiv \frac{e}{\pi} \theta$.

The quantum Hamiltonian invariant under the topologically trivial
gauge transformations is not unique. We can add to it any linear
combination of the gauge-invariant operators  $\rho(\pm p)$:
\[
\hat{\tilde{\rm H}} \rightarrow \hat{\tilde{\rm H}} +
\beta_{\rm H,0} + \sum_{p>0} ( \beta_{\rm H,+}(p) \rho(p) +
\beta_{\rm H,-}(p) \rho(-p) )
\]
where $\beta$$_{\rm H,0}$, $\beta$$_{\rm H,\pm}$ are yet undetermined
functions. The conditions ~\ref{eq: cetnol} does not clearly fix
these functions.

The Hamiltonian of the consistent quantum theory of the CSM
should be invariant under the topologically nontrivial gauge
transformations as well. So next to ~\ref{eq: cetnol} is the
following condition
\begin{equation}
[\hat{\tilde{\rm H}}_{\rm R} , \hat{\rm T}_b]_{-} =0.
\label{eq: cetvosem}
\end{equation}
The condition ~\ref{eq: cetvosem} can be  then rewritten as a
system of three equations for $(\beta_{\rm H,0},\beta_{\rm H,\pm})$
and is solved up to constants independent of $[\frac{eb\rm L}{2\pi}]$
by
\begin{eqnarray}
\beta_{\rm H,0}^{s} & = & \hbar ([\frac{eb\rm L}{2\pi}])^2
\sum_{p>0} \frac{1}{p} \varepsilon_{\rm R}^{s}(p), \nonumber \\
\beta_{\rm H,\pm}^{s} & = & \hbar [\frac{eb\rm L}{2\pi}]
\frac{(-1)^p}{p} \varepsilon_{\rm R}^{s}(p),
\label{eq: cetdevet}
\end{eqnarray}
where
\[
\varepsilon_{\rm R}^{s}(p)  \equiv  \frac{2\pi}{\rm L} p
|\lambda \varepsilon_{p,\rm R}|^{-s} +
\frac{e^2\rm L}{4{\pi}^2} \frac{1}{p} \hbar.
\]

$3$. If we apply the bosonization procedure, then the bosonized
version of the regularized free fermionic Hamiltonian is (see
Appendix )
\[
\hat{\rm H}_0^s = \frac{2\pi}{\rm L} \hbar \sum_{p>0}
|\lambda \varepsilon_{p,\rm R}|^{-s} {\rho}_{s}(-p) {\rho}_{s}(p)
\]
With ~\ref{eq: cetodin} and ~\ref{eq: cetdevet}, on the physical
states we then get
\[
\hat{\tilde{\rm H}}_{\rm R} |{\rm phys};A \rangle =
\hat{\rm H}_{\rm phys} |{\rm phys};A \rangle
\]
where
\[
\hat{\rm H}_{\rm phys}^s =
\frac{\hbar}{2} \sum_{p>0}  \frac{1}{p} {\varepsilon}_{\rm R}^{s}(p)
\rho_s(-p) \rho_s(p) + \hbar [\frac{eb\rm L}{2\pi}]
\sum_{p>0} \frac{(-1)^p}{p}{\varepsilon}_{\rm R}^{s}(p)
\rho_s(p)
\]
\[
+ \frac{1}{2{\rm L}^2} \int_{-{\rm L}/2}^{{\rm L}/2} dx \cdot
(\hat{\tilde{\pi}}_b^s(x))^2   - \frac{1}{2} \xi_{\rm R} \hbar.
\]

We have defined here the modified momentum operator
\[
\hat{\tilde{\pi}}_{b}^{s}(x) = \hat{\pi}_b - \frac{\rm L}{2}
\hbar {\cal E}_{s}(x)
\]
where
\[
{\cal E}_{s}(x) \equiv - \frac{2{e_s}x}{\rm L}
[\frac{eb\rm L}{2\pi}]
\]
and
\begin{equation}
e_s \equiv \sqrt{e^2 + \frac{48\pi}{{\rm L}^2} \frac{1}{\hbar}
\sum_{p>o} |\lambda \varepsilon_{p,\rm R}|^{-s}} .
\label{eq: petnol}
\end{equation}
We see that the line integral $b$ not only represents the physical
degrees of freedom of the gauge field, but also creates on the circle
$-\frac{\rm L}{2} \leq x \leq \frac{\rm L}{2}$ a background linearly
rising electric field in which the physical degrees of freedom of the
model are moving. On the states  ~\ref{eq: cetshest} the density
of the electromagnetic part of the physical Hamiltonian is
\[
\hat{\cal H}_{\rm EM}^{s,{\rm phys}} \rightarrow \frac{1}{2{\rm L}^2}
(\hat{\pi}_b + \hbar \frac{\rm L}{2} ({\cal E}_{\theta} -
{\cal E}_{s}))^2.
\]
While the constant background electric field is general in
one-dimensional gauge models defined on the circle, the linearly
rising one is specific to the CSM  \cite{sarad92}.

For large $\rm L$, we may neglect the second term in the
parentheses in ~\ref{eq: petnol}, so $e_s \simeq e$ and
\[
{\cal E}_{s}(x) \simeq - \frac{2ex}{\rm L} [\frac{eb\rm L}{2\pi}]
\]
that coincides with the expression given for the background electric
field strength in  \cite{sarad92}. If we evaluate $e_s$ at large
$s$ and then take the limit $s \to 0$, we get again that
\[
\lim_{s \to 0} e_s = e
\]
and
\[
\lim_{s \to 0} {\cal E}_{s}(x) = - \frac{2ex}{{\rm L}}
[\frac{eb{\rm L}}{2\pi}].
\]
The commutation relations for $\hat{\tilde{\pi}}_b$ are
\[
[\hat{\tilde{\pi}}_{b}(x) , \hat{\tilde{\pi}}_{b}(y) ]_{-} =
i {\hbar}^2 \frac{e^2{\rm L}}{2\pi} (x-y).
\]
The background linearly rising electric field may be described by
the scalar potential
\[  \varphi(x) = \frac{e}{\rm L} x^2 [\frac{eb\rm L}{2\pi}]  \]
and is created by the charge uniformly distributed on the circle
with the density
\[
{\rho}(x) = - \frac{2}{\rm L} [\frac{eb\rm L}{2\pi}].
\]
The topologically nontrivial gauge transformations change $\rho(x)$
as follows
\[ {\rho(x)} \rightarrow {\rho(x)} - \frac{2}{\rm L} n.   \]
Only non-zero $[\frac{eb\rm L}{2\pi}]$'s correspond to the nonvanishing
background charge density. Moreover, for non-zero $[\frac{eb\rm L}
{2\pi}]$ the fermionic physical degrees of freedom and $b$ are not
decoupled in the physical Hamiltonian. Such decoupling for all
values of $[\frac{eb\rm L}{2\pi}]$ is known to occur in the Schwinger
model  \cite{hetr88,iso90}. It is just the background linearly
rising electric field that couples $b$ to the fermionic physical degrees
of freedom.

We see also that the spectrum of the fermionic part of the physical
Hamiltonian
\[ {\varepsilon}_{\rm R}(p) = \lim_{s \to \infty} {\varepsilon}_{\rm R}
^{s}(p) = \frac{\rm L}{2{\pi}p} [ (\frac{2{\pi}p}{\rm L})^2 +
\frac{e^2}{2\pi} \hbar]       \]
is non-relativistic that indicates the breakdown of relativistic
invariance.

\subsection{EXOTIZATION}

Let us present now the procedure which we call exotization. We can
formally decouple the matter and gauge field degrees of freedom by
introducing the exotic statistics matter field  \cite{sarad94}.

We define the composite field
\begin{equation}
\tilde{\psi}_{\rm R}(x) = \exp\{i \frac{\pi}{\rm L} x +
\frac{i}{\hbar} \frac{2\pi}{e\rm L} \hat{\tilde{\pi}}_{b}(x)\}
\cdot {\psi}_{\rm R}(x) .
\label{eq: petodin}
\end{equation}
The field $\tilde{\psi}_{\rm R}(x)$ has the commutation relations
\begin{eqnarray}
\tilde{\psi}_{{\rm R}}^\star (x) \tilde{\psi}_{{\rm R}}(y) & + &
e^{+i{\rm F}(x,y)}  \tilde{\psi}_{\rm R}(y)
\tilde{\psi}_{\rm R}^{\star}(x)  =  \delta (x-y), \nonumber \\
\vspace{5 mm}
\tilde{\psi}_{\rm R}(x)  \tilde{\psi}_{\rm R}(y) & + &
e^{-i{\rm F}(x,y)}  \tilde{\psi}_{\rm R}(y)  \tilde{\psi}_{\rm R}
(x)  = 0,
\label{eq: petdva}
\end{eqnarray}
where ${\rm F}(x,y) \equiv \frac{2\pi}{\rm L}(x-y)$ .
The commutation relations ~\ref{eq: petdva} are indicative of an
exotic statistics of $\tilde{\psi}_{\rm R}(x)$. This field is neither
fermionic nor bosonic. Only for $x=y$ Eqs.~\ref{eq: petdva} become
anti-commutators: $\tilde{\psi}_{\rm R}(x)$ ( and $\tilde{\psi}_{\rm R}
^{\star}(x)$ ) anticommutes with itself, i.e. behaves as a fermionic
field.

Using ~\ref{eq: shest} and ~\ref{eq: petodin}, we get the Fourier
expansion for the exotic field $\tilde{\psi}_{\rm R}(x)$ :
\[
\tilde{\psi}_{\rm R}^{s}(x) = \sum_{n \in \cal Z} \tilde{a}_n
\langle x|n;{\rm R} \rangle |\lambda \varepsilon_{n,\rm R}|^{-s/2}
\]
where
\[
\tilde{a}_n  \equiv  \exp\{\frac{i}{\hbar} \frac{2\pi}{e\rm L}
\hat{\pi}_b\} a_{n-[\frac{eb\rm L}{2\pi}]} ,
\]
\[
\tilde{a}_n^{\dagger}  \equiv  a_{n-[\frac{eb\rm L}{2\pi}]}^{\dagger}
\exp\{-\frac{i}{\hbar} \frac{2\pi}{e\rm L} \hat{\pi}_b\} \neq \tilde{a}_{-n}.
\]
The exotic creation and annihilation operators $\tilde{a}_n ,
\tilde{a}_n^{\dagger}$ fulfil the following commutation relations
algebra:
\begin{eqnarray*}
\tilde{a}_n^{\dagger} \tilde{a}_m   +  \tilde{a}_{m-1}
\tilde{a}_{n-1}^{\dagger} & = & \delta_{m,n}, \\
\tilde{a}_n  \tilde{a}_m^{\dagger}  +  \tilde{a}_{m+1}^{\dagger}
\tilde{a}_{n-1} & = & \delta_{m,n} ,  \\
\tilde{a}_n  \tilde{a}_m  + \tilde{a}_{m+1} \tilde{a}_{n-1} & = & 0,\\
\tilde{a}_n^{\dagger}  \tilde{a}_m^{\dagger}  +
\tilde{a}_{m-1}^{\dagger}  \tilde{a}_{n+1}^{\dagger} & = & 0 .
\end{eqnarray*}
Let us introduce the new Fock vacuum $\overline{|{\rm vac};A \rangle}$
defined as
\begin{eqnarray*}
\tilde{a}_n \overline{|{\rm vac};A \rangle}=0
& {\rm for} & n>[\frac{eb\rm L}{2\pi}], \\
\tilde{a}_n^{\dagger} \overline{|{\rm vac};A \rangle}=0
& {\rm for} & n \leq [\frac{eb\rm L}{2\pi}] - 1
\end{eqnarray*}
and denote the normal ordering with respect to $\overline{|{\rm vac};A
\rangle}$ by $\vdots \hspace{2 mm} \vdots$ .
The exotic matter current operator is
\[
\hat{\tilde{j}}_{\rm R}^{s}(x)  =  \hbar \sum_{n \in \cal Z}
\frac{1}{\rm L} \exp\{i \frac{2\pi}{\rm L} nx\} \cdot \tilde{\rho}_{s}(n),
\]
\[
\tilde{\rho}_{s}(n)  =  \sum_{k \in \cal Z} \frac{1}{2}
[\tilde{a}_{k}^{\dagger} , \tilde{a}_{k+n}]_{-} \cdot
|\lambda \varepsilon_{k, \rm R}|^{-s} |\lambda \varepsilon_{k+n,\rm R}|
^{-s}.
\]
The new operators $\tilde{\rho}(n)$ and the old ones $\rho(n)$
are connected in the following way:
\[
\tilde{\rho}(n) = \rho(n) + \delta_{n,0} [\frac{eb\rm L}{2\pi}].
\]
The total exotic matter charge is
\[
\hat{\tilde{\rm Q}}_{\rm R} = \hat{\rm Q}_{\rm R} + \hbar
[\frac{eb\rm L}{2\pi}].
\]
On the physical states ~\ref{eq: dvavosem}
\begin{equation}
\hat{\tilde{\rm Q}}_{\rm R,\rm N} |{\rm phys};A\rangle \equiv
\hbar \vdots \tilde{\rho}(0) \vdots |{\rm phys};A\rangle =
\hbar [\frac{eb\rm L}{2\pi}] |{\rm phys};A \rangle.
\label{eq: pettri}
\end{equation}
The old creation and annihilation operators act on the new Fock
vacuum by the rule:
\begin{eqnarray*}
a_n  \overline{|{\rm vac};A \rangle}  = 0  & {\rm for} & n>0,\\
a_n^{\dagger} \overline{|{\rm vac};A \rangle}=0 & {\rm for} & n \leq 0.
\end{eqnarray*}
If we compare the old and the new Fock vacuum states, then we see
a shift of the level that separates the filled levels and the empty
ones. The new Fock vacuum is defined such that the levels with energy
lower than (or equal to) the energy of the level $n=0$ are filled
and the others are empty, i.e. the background charge is incorporated
in the new Fock vacuum.

Using ~\ref{eq: pettri}, we rewrite $\hat{\rm H}_{\rm phys}$ in the
compact form with matter and gauge-field degrees of freedom decoupled:
\[
\hat{\rm H}_{\rm phys} = \hat{\rm H}_u + \hat{\rm H}_{\rm matter},
\]
where
\[
\hat{\rm H}_u \equiv {\rm L} \{ \frac{1}{2{\rm L}^2}
(\hat{\pi}_u + \hbar \frac{e\rm L}{2\pi} \theta)^2 +
\frac{e^2}{4\pi} \hbar u^2 \}
\]
is a Hamiltonian governing the dynamics of the fractional part
of $\frac{eb\rm L}{2\pi}$ :
\[
u \equiv \frac{2\pi}{e\rm L} ( \{\frac{eb\rm L}{2\pi} \} -
\frac{1}{2} ), \hspace{5 mm} \hat{\pi}_u \equiv -i\hbar \frac{d}{du},
\]
while  the matter Hamiltonian is
\[
\hat{\rm H}_{\rm matter} = \hat{\rm H}_{(1)}
+\hat{\rm H}_{(2)},
\]
\begin{equation}
\hat{\rm H}_{(1)} \equiv \frac{\hbar}{2}
\sum_{\stackrel{p \in \cal Z}{p \neq 0}}
\frac{1}{p} \varepsilon_{\rm R}(p) \rho_{tot}(-p) \rho_{tot}(p) ,
\label{eq: petcet}
\end{equation}
\[
\hat{\rm H}_{(2)} \equiv {\hbar}^2 \frac{e^2{\rm L}}{32{\pi}^2}
\sum_{\stackrel{p \in \cal Z}{p \neq 0}}
\sum_{\stackrel{q \in \cal Z}{q \neq 0}}
\frac{(-1)^{p+q}}{pq} \rho(-p) \rho(q) .
\]
The second term $\hat{\rm H}_{(2)}$ appears after
solving the constraint ~\ref{eq: cetsem}. The operators
\[
\rho_{tot}(\pm p) \equiv \rho(\pm p) + (-1)^p \frac{1}{\hbar}
\hat{\tilde{\rm Q}}_{\rm R,\rm N}
\]
are invariant under both topologically trivial and nontrivial
gauge transformations.

Thus, the physical quantum CSM can be formulated in two equivalent ways.
In the first way, the matter fields are fermionic and coupled nontrivially
to the global gauge-field degree of freedom. In the second way, the matter
and gauge-field degrees of freedom are decoupled in the Hamiltonian, but
the matter fields acquire exotic statistics. It is the background-matter
interaction that leads to exotic statistics of the matter fields. The
concepts of background-matter interaction and exotic statistics are
therefore closely related.
\newpage
\section{Poincare Algebra}
\label{sec: poinc}
$1$. The classical momentum and boost generators are given by
\[
{\rm P}_{\rm R}=\int dx(-i\hbar \psi^{\star}_{\rm R} \partial_1
\psi_{\rm R} - {\rm E} \partial_1 A),
\]
\[
{\rm K}_{\rm R} =  \int dx \cdot x {\cal H}_{\rm R} .
\]
The momentum generator translates $A_1$ and ${\psi}_{\rm R}$ in space:
\begin{eqnarray*}
\{ {\rm P}_{\rm R},{A_1}(y) \} & = & \partial_1 {A_1}(y) , \\
\{ {\rm P}_{\rm R},{\psi}_{\rm R}(y) \} & = & \partial_1
{\psi}_{\rm R}(y),
\end{eqnarray*}
while the boost generator acts as follows
\begin{eqnarray*}
\{ {\rm K}_{\rm R},{A_1}(y) \} & = & -y \dot{A}_1(y) ,\\
\{ {\rm K}_{\rm R},{\psi}_{\rm R}(y) \} & = & -y\dot{\psi}_{\rm R}(y).
\end{eqnarray*}
After a straightforward calculation we obtain the algebra
\[
\{ {\rm H}_{\rm R} , {\rm P}_{\rm R} \} = 0 ,
\]
\[
\{ {\rm P}_{\rm R} , {\rm K}_{\rm R} \} = - {\rm H}_{\rm R} ,
\hspace{5 mm}
\{ {\rm H}_{\rm R} , {\rm K}_{\rm R} \} = - {\rm P}_{\rm R},
\]
i.e. at the classical level, these generators obey the Poincare
algebra.

At the quantum level, the momentum generator becomes
\begin{equation}
\hat{\rm P}^s_{\rm R} = \frac{1}{2} \hbar \int dx
( {\psi}^{\dagger s}_{\rm R} (-i \partial_x) {\psi}^s_{\rm R}
- {\psi}^s_{\rm R} (i \partial_x) {\psi}^{\dagger s}_{\rm R} )
- \int dx \hat{\rm E} {\partial_1}A_1 .
\label{eq: petshest}
\end{equation}
Using the Fourier expansions ~\ref{eq: shest}, ~\ref{eq: odinshest}
and the quantum Gauss law constraint ~\ref{eq: dvasem}, we rewrite
the quantum momentum as
\begin{equation}
\hat{\rm P}^s_{\rm R} = \hat{\rm H}^s_0 - \frac{e^2 \rm L}{2 \pi} \hbar
\sum_{p > 0} {\alpha}_{p}{\alpha}_{-p} - \frac{1}{2} \xi_{\rm R} \hbar -
\frac{1}{2} eb{\rm L} \cdot \eta_{\rm R} \hbar .
\label{eq: petsem}
\end{equation}
As the Hamiltonian, the momentum generator is not unique. We act
in the same way as before in Section~\ref{sec: exoti} . To get
the physical momentum generator, we first define
\[
\hat{\tilde{\rm P}}_{\rm R} \equiv \hat{\rm P}_{\rm R} +
\frac{1}{2} \sum_{p>0} \{[\hat{v}_{\rm P,+} ,\hat{\rm G}_{+}]_{+}
+ [\hat{v}_{\rm P,-} , \hat{\rm G}_{-}]_{+} \}
\]
and impose the condition
\begin{equation}
[\hat{\tilde{\rm P}}_{\rm R}, \hat{\tilde{\rm G}}_{\pm}(p)]_{-}=0.
\label{eq: doptri}
\end{equation}
The condition ~\ref{eq: doptri} fix $\hat{v}_{\rm P,\pm}$ and makes
the momentum operator invariant under the topologically trivial
gauge transformations. Next, we modify $\hat{\tilde{\rm P}}_{\rm R}$
by
\begin{equation}
\hat{\tilde{\rm P}}_{\rm R} \rightarrow \hat{\tilde{\rm P}}_{\rm R}
+ \beta_{\rm P,0} +
\sum_{p > 0} ( \beta_{\rm P,+}(p) \rho(p) +
\beta_{\rm P,-}(p) \rho(-p))
\label{eq: dopcet}
\end{equation}
in order to make it invariant under the nontrivial gauge transformations
as well:
\begin{equation}
[ \hat{\tilde{\rm P}}_{\rm R} , \hat{\rm T}_b ]_{-}  =  0 .
\label{eq: doppet}
\end{equation}
Finding $\hat{v}_{\rm P,\pm}$ from Eq.~\ref{eq: doptri} and
$(\beta_{\rm P,0} , \beta_{\rm P,\pm})$ from Eq.~\ref{eq: doppet}
and substituting them  into ~\ref{eq: dopcet}, we get the physical
quantum momentum in the form
\[
\hat{\tilde{\rm P}}_{\rm R} |{\rm phys};A \rangle = \hat{\rm P}_{\rm phys}
|{\rm phys};A \rangle,
\]
\begin{equation}
\hat{\rm P}_{\rm phys} = \hat{\rm P}_{\rm matter} =
\frac{\pi}{\rm L} \hbar \sum_{\stackrel{p \in \cal Z}{p \neq 0}}
\rho_{\rm tot}(-p) \rho_{\rm tot}(p).
\label{eq: dopshest}
\end{equation}
The action of this operator on the matter fields $\rho(\pm p)$ is
\[
[\hat{\rm P}_{\rm phys} , \rho_{\rm tot}(\pm p)]_{-}  =
\mp \frac{2\pi}{\rm L} \hbar p \cdot \rho_{\rm tot}(\pm p).
\]
The quantum boost generator is
\[
\hat{\rm K}_{\rm R} = -i \hbar \frac{\rm L}{2\pi}
\sum_{p>0} \frac{(-1)^p}{p} ({\cal H}_{\rm R}(p) -{\cal H}_{\rm R}(-p)),
\]
where
\[
{\cal H}_{\rm R}(p) = {\cal H}_{\rm F}(p) + {\cal H}_{\rm EM}(p),
\]
${\cal H}_{\rm F}(p)$ and ${\cal H}_{\rm EM}(p)$ being given by
Eqs.~\ref{eq: dopdva} and ~\ref{eq: dopodin} correspondingly.

The physical quantum boost generator can be constructed in the same way
as the physical Hamiltonian and momentum and has the form
\[
\hat{\rm K}_{\rm phys}  =  \hat{\rm K}_{\rm matter}=
\hat{\rm K}_{(1)} + \hat{\rm K}_{(2)} ,
\]
\begin{equation}
\hat{\rm K}_{(1)} \equiv - \frac{i\hbar}{2}
\sum_{\stackrel{p \in \cal Z}{p \neq 0}} \frac{(-1)^p}{p}
\sum_{\stackrel{q \in \cal Z}{q \neq (0;-p)}} {\rm k}_{\rm R}(p,q)
\cdot \rho_{\rm tot}(-q) \rho_{\rm tot}(p+q),
\label{eq: dopsem}
\end{equation}
\[
\hat{\rm K}_{(2)}  =  \frac{\hbar}{8\pi} (\frac{e\rm L}{\pi})^2
\sum_{\stackrel{p \in \cal Z}{p \neq 0}} \frac{(-1)^p}{p^2}
\rho_{\rm tot}(p) \cdot \hat{\pi}_{[\frac{eb\rm L}{2\pi}]},
\]
where
\[
{\rm k}_{\rm R}(p,q) \equiv 1 + \frac{\hbar}{8\pi}
(\frac{e\rm L}{\pi})^2 \frac{1}{q(q+p)}.
\]
On the states ~\ref{eq: cetsem} $\hat{\rm K}_{(2)}$ becomes
\[
\hat{\rm K}_{(2)} = i \frac{{\hbar}^2}{16\pi} (\frac{e\rm L}{\pi})^2
\sum_{\stackrel{p \in \cal Z}{p \neq 0}} \frac{(-1)^p}{p^2}
\sum_{\stackrel{q \in \cal Z}{q \neq 0}} \frac{(-1)^q}{q}
\rho(-q) \rho_{\rm tot}(p) .
\]
$2$. Let us now construct the algebra of the physical Hamiltonian,
momentum and boost generators. Since the relativistic invariance
is broken, this algebra is not definetely a Poincare one. We neglect,
for the moment, the global gauge-field degree of freedom contribution
and start with the following generators :
\begin{eqnarray*}
\hat{\rm H}_{\rm phys} & = & \sum_{\stackrel{p \in \cal Z}{p \neq 0}}
{\rm H}(p) , \\
\hat{\rm P}_{\rm phys} & = & \sum_{\stackrel{p \in \cal Z}{p \neq 0}}
{\rm P}(p) , \\
\hat{\rm K}_{\rm phys} & = & \sum_{\stackrel{p \in \cal Z}{p \neq 0}}
\sum_{\stackrel{q \in \cal Z}{q \neq (0,-p)}} {\rm K}(p,q),
\end{eqnarray*}
where
\begin{eqnarray*}
{\rm H}(p) & \equiv & \frac{\hbar}{2} \frac{1}{p}
\varepsilon_{\rm R}(p) \rho(-p) \rho(p), \\
{\rm P}(p) & \equiv & \hbar \frac{\pi}{\rm L} \rho(-p) \rho(p), \\
{\rm K}(p,q) & \equiv & - \frac{i\hbar}{2} \frac{(-1)^p}{p}
{\rm k}_{\rm R}(p,q) \rho(-q) \rho(p+q).
\end{eqnarray*}
We can check by a straightforward calculation that
\[
[ \hat{\rm H}_{\rm phys} , \hat{\rm P}_{\rm phys} ]_{-}  =  0,
\]
i.e. the translational invariance is preserved , while two other
commutation relations
\[
[\hat{\rm P}_{\rm phys} , \hat{\rm K}_{\rm phys}]_{-}  \neq
-i\hbar \hat{\rm H}_{\rm phys},
\]
\[
[\hat{\rm H}_{\rm phys} , \hat{\rm K}_{\rm phys}]_{-}  \neq
-i\hbar \hat{\rm P}_{\rm phys}
\]
differ from those of Poincare algebra. In terms of ${\rm H}(p),
{\rm P}(p), {\rm K}(p,q)$ ($p , q$ are nonzero) these commutation
relations are written in a compact form as follows
\[
[{\rm H}(p) , {\rm K}(q,m)]_{-} = \frac{\hbar}{2} \varepsilon_{\rm R}(p)
\{ {\rm K}(q,p) \cdot (\delta_{m,-p-q}+\delta_{m,p})-(p \rightarrow -p) \},
\hspace{5 mm} q \neq \pm p,
\]
\[
[{\rm H}(p) , {\rm K}(\pm p,m)]_{-} = \pm \frac{\hbar}{2}
\varepsilon_{\rm R}(p) \cdot {\rm K}(\pm p, \pm p) (\delta_{m, \mp 2p} +
\delta_{m, \pm p})
\]
and
\[
[{\rm P}(p) , {\rm K}(q,m)]_{-} =  \hbar \frac{\pi}{\rm L} p
\{ {\rm K}(q,p) \cdot (\delta_{m,-p-q} + \delta_{m,p})
- (p \rightarrow -p)\} , \hspace{ 5mm} q \neq \pm p,
\]
\[
[{\rm P}(p) , {\rm K}(\pm p,m)]_{-} = \pm \hbar \frac{\pi}{\rm L}
p {\rm K}(\pm p, \pm p) \cdot (\delta_{m,\mp 2p} + \delta_{m,\pm p}).
\]
If we introduce
\[
b^{\pm}(p) \equiv \frac{1}{\sqrt{p}} \rho(\mp p) ,
\hspace{5 mm} p>0,
\]
then
\begin{eqnarray*}
\hat{\rm H}_{\rm phys} & = & \hbar \sum_{p>0} \varepsilon_{\rm R}(p)
b^{+}(p) b^{-}(p) ,\\
\hat{\rm P}_{\rm phys} & = & \hbar \sum_{p>0} \frac{2\pi}{\rm L} p
b^{+}(p) b^{-}(p) .
\end{eqnarray*}
Therefore, $b^{+}(p)$ and $b^{-}(p)$ can be interpreted respectively
as the creation and annihilation operators for a particle of momentum
$\hbar \frac{2\pi}{\rm L} p$ and energy $\hbar \varepsilon_{\rm R}(p)$.

If the global gauge-field degree of freedom contribution is taken into
account, then the translational invariance is also lost. Indeed,
the total matter Hamiltonian ~\ref{eq: petcet} is not invariant
under translations, since the second term
\[
\hat{\rm H}_{(2)} = \sum_{\stackrel{p \in \cal Z}{p \neq 0}}
\sum_{\stackrel{q \in \cal Z}{q \neq 0}} {\rm H}(p,q),
\]
\[
{\rm H}(p,q) \equiv {\hbar}^2 \frac{e^2 \rm L}{32{\pi}^2}
\frac{(-1)^{p+q}}{pq} \rho(-p) \rho(q)
\]
does not commute with the physical momentum:
\[
[{\rm H}(p,q),{\rm P}(m)]_{-}= \frac{{\hbar}^3e^2}{32\pi}
\frac{(-1)^{p+q}}{pq} \{ p \rho_(-q) \rho_{\rm tot}(p) \cdot
(\delta_{p,-m} + \delta_{p,m})
\]
\[
- q \rho_{\rm tot}(-q) \rho(p)
\cdot (\delta_{q,-m} + \delta_{q,m}) \}.
\]
All three commutators of the Poincare algebra are therefore broken.
The spectrum of the model is nonrelativistic, and there is no mass
in this spectrum.

$3$. In the limit $\rm L \to \infty$, when the model is defined
on the line ${\rm R}^1$ , $b$ vanishes and the gauge field
does not possess any physical degree of freedom.

The physical Hamiltonian and momentum commute. Two other commutation
relations of the Poincare algebra are broken. As before with
$\rm L$ finite , the reason for the breaking of the relativistic
invariance is anomaly or , more exactly, the fact that the local gauge
symmetry is realized projectively.

For $\rm L \to \infty$, we can however construct the states which
are simultaneous eigenstates of the Hamiltonian and momentum.
The corresponding eigenvalues are connected in a relativistic way
and allow us to interpret these states as massive \cite{sarad91}.
\newpage
\section{Charge Screening}
\label{sec: scree}
Let us introduce a pair of external charges, namely, a positive charge
with strength $q$ at $x_0$ and a negative one with the same strength
at $y_0$. The external current density is
\[
j_{\rm ex,0}(x) = q(\delta(x-x_0) - \delta(x-y_0)) =
\frac{1}{\rm L} \sum_{p \in \cal Z} j_p^{\rm ex}
\exp\{-i \frac{2\pi p}{\rm L}x\},
\]
where
\[
j_p^{\rm ex} \equiv q(e^{i2\pi px_0/{\rm L}} - e^{i2\pi py_0/{\rm L}}).
\]
The total external charge is zero, so the external current density
has vanishing zero mode, $j_0^{\rm ex}=0$.
The Lagrangian density of the CSM changes as follows
\[
{\cal L} \longrightarrow {\cal L} + e A_0 \cdot j_{\rm ex,0}.
\]
The classical CSM with the external charges added can be quantized
in the same way as that without external charges.
The quantum Gauss' law operator becomes
\[
\hat{\rm G}_{\rm ex} \equiv \hat{\rm G} + e j_{\rm ex,0} =
\partial_1 \hat{\rm E} + e (\hat{j}_{\rm R} + j_{\rm ex,0}).
\]
Its Fourier expansion is
\[
\hat{\rm G}_{\rm ex} = \hat{\rm G}_0 +
\frac{2\pi}{{\rm L}^2} \sum_{p>0} (\hat{\rm G}_{+}^{\rm ex}(p)
e^{i\frac{2\pi}{\rm L}px} - \hat{\rm G}_{-}^{\rm ex}(p)
e^{-i\frac{2\pi}{\rm L}px} ),
\]
where
\begin{eqnarray*}
\hat{\rm G}_{+}^{\rm ex}(p) & \equiv & \hat{\rm G}_{+}(p) +
\frac{e\rm L}{2\pi} (j_p^{\rm ex})^{\star} , \\
\hat{\rm G}_{-}^{\rm ex}(p) & \equiv & \hat{\rm G}_{-}(p) -
\frac{e\rm L}{2\pi} j_p^{\rm ex}.
\end{eqnarray*}
The physical states $|{\rm phys};A;{\rm ex} \rangle$ are defined as
\[
\hat{\tilde{\rm G}}_{\pm}^{\rm ex}(p) |{\rm phys};A;\rm ex \rangle
\equiv (\hat{\rm G}_{\pm}^{\rm ex}(p) \pm  \hbar \frac{e^2{\rm L}^2}
{8{\pi}^2} \alpha_{\pm p}) |{\rm phys};A;\rm ex \rangle =0.
\]
The external charges change also the Fock vacuum. We have the
following definition for the Fock vacuum in the presence of the
external charges:
\begin{eqnarray}
(\rho(p) + \frac{1}{\hbar} (j_p^{\rm ex})^{\star})
|{\rm vac};A;{\rm ex} \rangle & = & 0 , \nonumber \\
\langle {\rm ex}; {\rm vac};A| (\rho(-p) + \frac{1}{\hbar}
j_p^{\rm ex} ) & = & 0 , \hspace{5 mm} {\rm for} \hspace{5 mm} p>0 .
\label{eq: dopdd}
\end{eqnarray}
The physical quantum matter Hamiltonian invariant under
the both topologically trivial and nontrivial gauge transformations
becomes
\[
\hat{\rm H}_{(1)} =
\frac{\pi}{\rm L} \hbar \sum_{\stackrel{p \in \cal Z}{p \neq 0}}
\rho_{\rm tot}(-p) \rho_{\rm tot}(p)
+ \frac{e^2 \rm L}{8{\pi}^2} {\hbar}^2
\sum_{\stackrel{p \in \cal Z}{p \neq 0}}  \frac{1}{p^2}
(\rho(-p) + \frac{1}{\hbar} j_p^{\rm ex}) \cdot
(\rho(p)  + \frac{1}{\hbar} (j_p^{\rm ex})^{\star}),
\]
$\hat{\rm H}_{\rm matter,(2)}^{\rm phys}$ being given again by
Eq.~\ref{eq: petcet}.

We consider two different cases. $1$. Let us neglect the global
gauge-field degree of freedom contribution to the matter Hamiltonian.
After some calculations we rewrite it as
\begin{equation}
\hat{\rm H}_{\rm matter} = \frac{\hbar}{2}
\sum_{\stackrel{p \in \cal Z}{p \neq 0}} \frac{1}{p}
\varepsilon_{\rm R}(p) \rho_{\rm ex}(-p) \rho_{\rm ex}(p)
+ \frac{e^2}{2\pi} \sum_{p>0} \frac{1}{p\varepsilon_{\rm R}(p)}
(j_p^{\rm ex})^{\star}(j_p^{\rm ex}),
\label{eq: shestvosem}
\end{equation}
where
\begin{eqnarray}
\rho_{\rm ex}(p) & \equiv  & \rho(p) + \frac{e^2\rm L}{4{\pi}^2}
\frac{1}{p\varepsilon_{\rm R}(p)} (j_p^{\rm ex})^{\star}, \nonumber \\
\rho_{\rm ex}(-p) & \equiv  & \rho(-p) + \frac{e^2\rm L}{4{\pi}^2}
\frac{1}{p\varepsilon_{\rm R}(p)} j_p^{\rm ex}.
\label{eq: dopdp}
\end{eqnarray}
The ground state of this Hamiltonian differs from the vacuum one
{}~\ref{eq: dopdd} and is defined as
\begin{eqnarray*}
\rho_{\rm ex}(p) |{\rm ground};{\rm ex} \rangle & = & 0, \\
\langle {\rm ex}; {\rm ground}| \rho_{\rm ex}(-p) & = & 0,
\hspace{5 mm} p>0. \\
\end{eqnarray*}
The first term in ~\ref{eq: shestvosem} is normal ordered with
respect to this state and the second one is its energy:
\[
{\rm E}_0 = \langle {\rm ground};\rm ex| \hat{\rm H}_
{\rm matter} |{\rm ground};\rm ex \rangle =
\frac{e^2}{2\pi} \sum_{p>0} \frac{1}{p\varepsilon_{\rm R}(p)}
(j_p^{\rm ex})^{\star} (j_p^{\rm ex}).
\]
The energy ${\rm E}_0$ depends only on the distance between the
external charges:
\[
{\rm E}_0 = 4 \frac{(eq)^2}{\rm L} \sum_{p>0}
\frac{1}{(\frac{2\pi p}{\rm L})^2 + \frac{e^2}{2\pi}\hbar}
\{1 - \cos(\frac{2\pi p}{\rm L}(x_0 -y_0))\}
\]
\[
=\frac{(eq)^2}{m_0} \frac{\cosh{\frac{\rm L m_0}{2}} -
\cosh(\frac{\rm L m_0}{2} - m_0|x_0 -y_0|)}{\sinh{\frac{\rm L m_0}
{2}}},
\]
where $m_0^2=\frac{e^2}{2\pi}\hbar$.
In the limit $\rm L \gg 1$, we get
\[
{\rm E}_0 = \frac{(eq)^2}{m_0} (1 - e^{-m_0|x_0 - y_0|}),
\]
i.e. the Yukawa potential.

The current density induced by the two external charges is
\[
\langle {\rm ground};\rm ex|\hat{j}_R(x)|{\rm ground};\rm ex \rangle
\equiv f(x,x_0) - f(x,y_0),
\]
where
\[
f(x,x_0) \equiv - \frac{e^2q}{2{\pi}^2} \hbar \sum_{p>0}
\frac{1}{p\varepsilon_{\rm R}(p)} \cos(\frac{2\pi p}{\rm L}(x-x_0)) =
- \frac{qm_0}{2} \frac{\cosh(\frac{\rm L m_0}{2} - m_0|x-x_0|)}
{\sinh{\frac{\rm L m_0}{2}}}.
\]
The induced current density is a sum of the current densities
induced by the each charge. In the limit $\rm L \gg 1$,
\[
f(x,x_0) \simeq - \frac{qm_0}{2} e^{-m_0|x-x_0|}
\]
and damps exponentially as $x$ goes far from $x_0$, so the external
charges are screened globally. If we are far away from the external
charges , we can not find them.

$2$. Let us now take into account the gauge-field contribution
and consider the total matter Hamiltonian . We can diagonalize it
in the following form
\[
\hat{\rm H}_{\rm matter} = \hbar \sum_{\stackrel{p \in \cal Z}{p \neq 0}}
\varepsilon_{\rm R}(p) \tilde{\rho}_{\rm ex}(-p) \tilde{\rho}_{\rm ex}(p)
+{\hbar}^2 \frac{e^2 \rm L}{32{\pi}^2}
\sum_{\stackrel{p \in \cal Z}{p \neq 0}}
\sum_{\stackrel{q \in \cal Z}{q \neq 0}}
\frac{(-1)^{p+q}}{pq} \tilde{\rho}_{\rm ex}(-p) \tilde{\rho}_{\rm ex}(q)
+ \tilde{\rm E}_0(x_0,y_0;b) ,
\]
where
\[
\tilde{\rho}_{\rm ex}(\pm p) \equiv \rho_{\rm ex}(\pm p)
+ (-1)^p [\frac{eb\rm L}{2\pi}]  \pm
iq \hbar \frac{e^4 \rm L}{16{\pi}^3}
\frac{(-1)^p}{\varepsilon_{\rm R}(p)} \cdot
\frac{d_2}{1+ \frac{e^2}{4\pi} \hbar d_1} ,
\]
with $\rho_{\rm ex}(\pm p)$ given by Eq.~\ref{eq: dopdp},
and
\[
d_1 \equiv \frac{\rm L}{4m_0} \cdot \frac{\cosh{\frac{{\rm L}m_0}{2}}
-\cosh{\frac{{\rm L}m_0}{4}}}{\sinh{\frac{{\rm L}m_0}{2}}} ,
\]
\[
d_2 \equiv \frac{\pi}{2m_0} \cdot \frac{m_0(\sinh{m_0x_0} -
\sinh{m_0y_0}) + (x_0 - y_0) \cosh{\frac{{\rm L}m_0}{4}}}
{\sinh{\frac{{\rm L}m_0}{2}}} .
\]
The ground state of the total matter Hamiltonian satisfies
\begin{eqnarray*}
\tilde{\rho}_{\rm ex}(p) \overline{|{\rm ground};{\rm ex}\rangle} & = & 0,\\
\overline{\langle{\rm ex};{\rm ground}|} \tilde{\rho}_{\rm ex}(-p)& = & 0,
\hspace{3 mm} {\rm for} \hspace{3 mm} p>0.
\end{eqnarray*}
The energy of the ground state is
\[
\tilde{\rm E}_0(x_0,y_0;b) =
\overline{\langle {\rm ground};\rm ex|} \hat{\rm H}_{\rm matter}
\overline{|{\rm ground};\rm ex \rangle}
\]
\[
={\rm E}_0(x_0 - y_0) + \hbar q \frac{e^2}{2\rm L} [\frac{eb\rm L}
{2\pi}] (x_0^2 - y_0^2) + {\hbar}^2 q^2 \frac{e^6\rm L}{32{\pi}^4}
\cdot \frac{d_2^2}{1 + \frac{e^2}{4\pi} \hbar d_1}
\]
(up to constants independent of $x_0$ and $y_0$).
In contrast with ${\rm E}_0(x_0 -y_0)$, this energy depends not only
on the distance between the external charges, but also separately
on $x_0$ and $y_0$.
For $\rm L \gg 1$, we have
\[
\tilde{\rm E}_0(x_0,y_0;b) \simeq \frac{(eq)^2}{m_0}
(1 - e^{-m_0 |x_0 - y_0|}) + \hbar q \frac{e^2}{2\rm L}
[\frac{eb\rm L}{2\pi}] (x_0^2 - y_0^2) +
\hbar q^2 \frac{e^4}{8\pi m_0} (x_0 - y_0)^2.
\]
The induced current density is
\begin{equation}
\overline{\langle {\rm ground};\rm ex |} \hat{j}_R(x)
\overline{| {\rm ground};\rm ex \rangle} =
\tilde{f}(x;x_0) - \tilde{f}(x;y_0) - \frac{1}{\rm L} \hbar
[\frac{eb\rm L}{2\pi}] ,
\label{eq: semtri}
\end{equation}
where
\[
\tilde{f}(x;x_0) = f(x;x_0) - \frac{e^4 q \rm L}{64\pi}
\frac{1}{m_0} {\hbar}^2 \frac{\sinh{m_0x}}{(\sinh{\frac{{\rm L}m_0}{2}})^2}
(m_0\sinh{m_0x_0} + x_0\cosh{\frac{{\rm L}m_0}{4}}).
\]
The last term in ~\ref{eq: semtri} is the same for all values of
$x$ and induced by the global gauge-field degree of freedom.
In the limit $\rm L \gg 1$, we have
\[
\tilde{f}(x,x_0) \simeq - \frac{q}{2} (m_0 e^{-m_0|x - x_0|} +
\hbar \frac{e^2}{2\pi} x_0 e^{-\frac{3}{4}\rm L m_0}
\sinh{m_0 x} ).
\]
The second term here is very small for large, but finite $\rm L$.
At the same time, it increases exponentially when $x$ goes to
infinity. The external charges are not therefore screened even
globally.
\newpage
\section{Discussion}
\label{sec: discu}
We have shown that the anomaly influences essentially the physical
quantum picture of the CSM. For the model defined on $S^1$, when
the gauge field has a global physical degree of freedom, the left--
right asymmetric matter content results in the background linearly
rising electric field or ,equivalently, in the exotic statistics
of the physical matter field. This is a new physical effect caused
just by the anomaly and absent in the standard Schwinger model.

The anomaly leads also to the breaking of the relativistic invariance.
We have constructed the Poincare generators and shown that their
algebra is not a Poincare one. The spectrum of the physical
Hamiltonian is not relativistic and does not contain a massive boson.

Next, the external charges are not screened. Owing to the global
gauge-field degree of freedom contribution to the physical Hamiltonian,
the current density induced by the external charges doers not vanish
globally. Thus, such phenomena as the dynamical mass generation
and the total screening of charges characteristic for the Schwinger
model do not take place for the CSM on $S^1$.

For the CSM defined on $R^1$, the physical quantum picture differs
from that on $S^1$. The gauge field has not any physical degree of
freedom, and the background electric field disappears. The current
density induced by the external charges damps exponentially far away
from them. The external charges are then globally screened.

The anomaly manifests itself only in the breaking of the relativistic
invariance. However, the theory is invariant under space translations.
As shown in  \cite{sarad91}, \cite{sarad93}, this allows us to
construct the massive states which are simultaneous eigenstates
of the physical Hamiltonian and momentum. The screening of the
external charges and the dynamical mass generation (although in a
different way) are therefore valid for the physical quantum CSM
on $R^1$.

\vspace{3 cm}

{\Large \bf Acknowledgement}

\vspace{1 cm}

The author thanks \"{O}mer Faruk Dayi for hospitality at TUBITAK.
\newpage
\begin{appendix}

{\Large \bf Appendix}

\vspace{1 cm}

i) In this appendix we prove the commutation relations
{}~\ref{eq: odinsem} -- ~\ref{eq: dvanol}. We start with the
commutation relation ~\ref{eq: odinsem}. It can be established
in different ways \cite{mant85,iso90}. Here we derive it using
the $\zeta$--function regularization scheme. For the regulated
operators $\rho_{s}(m)$ we get
\[
[\rho_{s}(m) , \rho_{s}(n)]_{-} = \sum_{k \in \cal Z} \frac{1}{2}
[a_k^{\dagger},a_{k+m+n}]_{-} \cdot |\lambda \varepsilon_{k, \rm R}|^{-s/2}
|\lambda \varepsilon_{k+m+n, \rm R}|^{-s/2}
\]
\[
\cdot ( |\lambda \varepsilon_{k+m, \rm R}|^{-s} -
|\lambda \varepsilon_{k+n, \rm R}|^{-s} ).
\]
Since the commutator $[\rho(m) , \rho(n)]_{-}$ is a C--number, we
calculate it by taking the corresponding vacuum expectation value:
\[
\langle {\rm vac};A| [\rho_{s}(m) , \rho_{s}(n)]_{-} |{\rm vac};A \rangle
= \delta_{m, -n} {\rm I}_{1}^{s}(m),
\]
where
\[
{\rm I}_{1}^{s}(m) \equiv - \frac{1}{2} \sum_{k \in \cal Z}
{\rm sign}(\varepsilon_k) |\lambda \varepsilon_{k, \rm R}|^{-s}
( |\lambda \varepsilon_{k+m, \rm R}|^{-s} -
|\lambda \varepsilon_{k-m, \rm R}|^{-s} ).
\]
We see that the commutator is nonvanishing only for $m=-n$. The
sum ${\rm I}_{1}^{s}(m)$ can be easily evaluated. In particular,
for $m>0$, we have
\[
\sum_{k \in \cal Z}  {\rm sign}(\varepsilon_k)
|\lambda \varepsilon_{k, \rm R}|^{-s} |\lambda \varepsilon_{k \pm m,
\rm R}|^{-s}
\]
\[
= \sum_{k>0}  \frac{1}{(k- \{ \frac{eb\rm L}{2\pi} \} )^s
(k - \{ \frac{eb\rm L}{2\pi} \} + m)^s }
- \sum_{k \geq 0}  \frac{1}{(k + \{ \frac{eb\rm L}{2\pi} \})^s
(k + \{ \frac{eb\rm L}{2\pi} \})^s }  \mp  m,
\]
so ${\rm I}_{1}^{s}(m) = m$ for all values of $s$ and
\[
[\rho(m) , \rho(n)]_{-} = \lim_{s \to 0} [\rho_{s}(m) ,
\rho_{s}(n)]_{-} = m \delta_{m, -n} .
\]
ii) Let us now calculate the derivatives $\frac{d}{db}\rho(m)$
and $\frac{d}{d{\alpha}_{\pm p}}\rho(m)$. With Eq.~\ref{eq: devet},
we have
\[
\frac{d}{db} \rho_{s}(m) = \frac{1}{2} \sum_{n \in \cal Z}
\sum_{k \in \cal Z} ( \langle n;{\rm R}|\frac{d}{db}|k; {\rm R} \rangle
\cdot [a_n^{\dagger} , a_{k+m}]_{-}  - \langle k+m; {\rm R}|\frac{d}{db}|
n; {\rm R} \rangle \cdot [a_k^{\dagger} , a_n]_{-} )
\]
\[
\cdot |\lambda \varepsilon_{k, \rm R}|^{-s/2}
|\lambda \varepsilon_{n, \rm R}|^{-s/2}
|\lambda \varepsilon_{k+m, \rm R}|^{-s/2}
\]
and
\[
\frac{d}{d{\alpha}_{\pm p}} \rho_{s}(m) = \frac{1}{2} \sum_{n \in \cal Z}
\sum_{k \in \cal Z} ( \langle n;\rm R |\frac{d}{d{\alpha}_{\pm p}}|
k; \rm R \rangle \cdot [a_n^{\dagger} , a_{k+m}]^{-} -
\langle k+m;\rm R|\frac{d}{d{\alpha}_{\pm p}}|n; \rm R \rangle
\cdot [a_k^{\dagger} , a_n]_{-} )
\]
\[
\cdot |\lambda \varepsilon_{k, \rm R}|^{-s/2}
|\lambda \varepsilon_{n, \rm R}|^{-s/2}
|\lambda \varepsilon_{k+m, \rm R}|^{-s/2}.
\]
Substituting
\[
\langle n; \rm R|  \frac{d}{db}  | k; \rm R \rangle  =
\frac{ie}{2} \rm L \delta_{n,k},
\]
\[
\langle n; \rm R|  \frac{d}{d{\alpha}_{\pm p}}  | k;\rm R \rangle  =
\pm \frac{e\rm L}{2{\pi}p} \delta_{k-n,\mp p}
\mp \frac{e\rm L}{2\pi} \frac{(-1)^p}{p} \delta_{n,k}
\]
into these equations and taking again the vacuum expactation values ,
we obtain
\[
\langle {\rm vac};A |  \frac{d}{db} \rho_{s}(m)  |{\rm vac};A \rangle
=  0,
\]
\[
\langle {\rm vac};A |  \frac{d}{d{\alpha}_{\pm p}} \rho_{s}(m) |
{\rm vac};A \rangle  =  - \frac{e\rm L}{4{\pi}p}
\delta_{m,\pm p}  \cdot {\rm I}_{2}^{s}(m),
\]
where
\[
{\rm I}_{2}^{s}(m) \equiv \sum_{k \in \cal Z} {\rm sign}(\varepsilon_
{k,\rm R}) |\lambda \varepsilon_{k,\rm R}|^{-s}
(|\lambda \varepsilon_{k-p,\rm R}|^{-s/2} -
|\lambda \varepsilon_{k+p, \rm R}|^{-s/2} ).
\]
For large $s$, ${\rm I}_{2}^{s}(m) \simeq 2{\rm I}_{1}^{s}(m) =2m$,
so we finally come to the Eqs.~\ref{eq: dvanol}.

iii) To prove ~\ref{eq: odinvosem}, we calculate the commutator
of the corresponding regulated operators:
\[
[\hat{\rm H}_0^s , \rho_s(p)]_{-} =
\frac{2\pi}{\rm L} \hbar \sum_{k \in \cal Z} \frac{k}{2}
[a_k^{\dagger},a_{k+p}]_{-} \cdot |\lambda\varepsilon_{k,\rm R}|^{-{3s}/2}
|\lambda \varepsilon_{k+p,\rm R}|^{-s/2}
\]
\[
- \frac{2\pi}{\rm L} \hbar \sum_{k \in \cal Z} \frac{k}{2}
[a_{k-p}^{\dagger},a_k]_{-} \cdot |\lambda\varepsilon_{k,\rm R}|^{-{3s}/2}
|\lambda \varepsilon_{k-p,\rm R}|^{-s/2}, \hspace{5 mm}  (p>0).
\]
If we make the redefinition $k-p \rightarrow k$ in the second sum,
then
\[
[\hat{\rm H}_0^s , \rho_s(p) ]_{-} =
\frac{2\pi}{\rm L} \hbar \sum_{k \in \cal Z} \frac{k}{2}
[a_k^{\dagger},a_{k+p}]_{-} \cdot |\lambda\varepsilon_{k,\rm R}|^{-s/2}
|\lambda \varepsilon_{k+p,\rm R}|^{-s/2}
(|\lambda \varepsilon_{k,\rm R}|^{-s} -
|\lambda \varepsilon_{k+p, \rm R}|^{-s})
\]
\[
- \frac{2\pi}{\rm L} \hbar p \sum_{k \in \cal Z} \frac{1}{2}
[a_k^{\dagger},a_{k+p}]_{-} \cdot |\lambda\varepsilon_{k,\rm R}|^{-s/2}
|\lambda \varepsilon_{k+p,\rm R}|^{-{3s}/2}.
\]
For large $s$, the first term vanishes, so
\[
[\hat{\rm H}_0^s , \rho_s(p)]_{-} = - \frac{2\pi}{\rm L} \hbar
p \rho_{s}(p) |\lambda \varepsilon_{p,\rm R}|^{-s}.
\]
In the limit $s \to 0$, we then get ~\ref{eq: odinvosem}.
Similarly, for $\rho(-p)$ we have
\[
[\hat{\rm H}_0^s , \rho_s(-p)]_{-} = \frac{2\pi}{\rm L} \hbar
p \rho_s(-p) |\lambda \varepsilon_{-p,\rm R}|^{-s}.
\]
It can be checked that the bosonized form of $\hat{\rm H}_0^s$ which
reproduces the last two equations is
\[
\hat{\rm H}_0^s = \frac{2\pi}{\rm L} \hbar \sum_{p>0}
|\lambda \varepsilon_{p,\rm R}|^{-s} \rho_s(p) \rho_s(-p).
\]
\end{appendix}


\newpage
\begin{thebibliography}{44}
\bibitem{schw63} J. Schwinger, Phys.Rev. D {\bf 128},2425 (1962);\\
in {\it Theoretical Physics, Trieste Lectures}, 1962 (IAEA, Vienna,
1963 ).
\bibitem{jack85} R. Jackiw and R. Rajaraman , Phys.Rev.Lett.
{\bf 54}, 1219 (1985).
\bibitem{raja85} R. Rajaraman, Phys.Lett. {\bf B154}, 305 (1985).
\bibitem{wign39} E. Wigher, Ann.Math. {\bf 40}, 149 (1939).
\bibitem{jack83} R. Jackiw, in {\it Relativity, Groups and
Topology II} \\
( Les Houches Summer School 1983) (North--Holland,
Amsterdam, 1984).
\bibitem{nels85} P. Nelson and L. Alvarez-Gaume, Commun.Math.Phys.
{\bf 99}, 103 (1985).
\bibitem{fadd86} L. D. Faddeev and S. L. Shatashvili, Phys.Lett.
{\bf B167}, 225 (1986).
\bibitem{fadd84} L. D. Faddeev, Phys.Lett. {\bf B14}, 81, (1984); \\
Nuffield Workshop on Supersymmetry and Supergravity, 1985.
\bibitem{hall86} I. G. Halliday, E. Rabinovici and A. Schwimmer,
Nucl.Phys.{\bf B268}, 413 (1986).
\bibitem{para88} M. B. Paranjape, Nucl.Phys.{\bf B307}, 649 (1988).
\bibitem{sarad91} F. M. Saradzhev, Int.J.Mod.Phys.{\bf A6},
3823 (1991).
\bibitem{sarad93} F. M. Saradzhev, Int.J.Mod.Phys.  {\bf A8},
2915, 2937 (1993).
\bibitem{niemi85} A. Niemi and G. Semenoff, Phys.Rev.Lett.
{\bf 55}, 927 (1985); {\bf 56}, 1019 (1986).
\bibitem{niemi86} A. Niemi and G. Semenoff, Phys.Lett. {\bf B175},
439 (1986).
\bibitem{semen87} G. Semenoff, in {\it Super Field Theory}, H. Lee
et al, eds. ( Plenum, NY, 1987).
\bibitem{sarad92} F. M. Saradzhev, Phys.Lett. {\bf B278}, 449
(1992).
\bibitem{sarad94} F. M. Saradzhev, Phys.Lett. {\bf B324}, 192
(1994).
\bibitem{dirac64} P. A. M. Dirac, {\it Lectures on Quantum Mechanics}
( Yeshiva Univ., NY , 1964).
\bibitem{mant85} N. S. Manton, Ann. Phys. {\bf 159}, 220 (1985).
\bibitem{raje88} S. Rajeev, Phys.Lett. {\bf B212}, 203 (1988).
\bibitem{hetr88} J. E. Hetrick and Y. Hosotani, Phys.Rev.
D {\bf 38}, 2621 (1988); \\
Phys.Lett. {\bf B230}, 88 (1989).
\bibitem{sarad88} F. M. Saradzhev, Sov.Phys. -- Lebedev Inst.
Reports, n. 9, 57 (1988); \\
Int.J.Mod.Phys. {\bf A9}, 1994, 3179 (1994).
\bibitem{niese86} A. Niemi and G. Semenoff, Phys.Repts. {\bf 135},
99 (1986).
\bibitem{schiff68} L. I. Schiff, {\it Quantum Mechanics} (McGraw--
Hill, NY , 1968).
\bibitem{berry84} M. V. Berry, Proc.R.Soc. London, {\bf A392},
45 (1984).
\bibitem{jack76} R. Jackiw and C. Rebbi, Phys.Rev.Lett.
{\bf 37}, 172 (1976).
\bibitem{callan76} C. G. Callan, Jr., R. Dashen and D. J. Gross,
Phys.Lett. {\bf B63}, 334 (1976).
\bibitem{iso90} S. Iso and H. Murayama, Progr.Theor.Phys.
{\bf 84}, 142 (1990).
\end{thebibliography}
\end{document}